\documentclass[11pt,a4paper]{article}

\usepackage[utf8]{inputenc}
\usepackage[T1]{fontenc}
\usepackage[margin=1in]{geometry}
\usepackage{amsmath,amssymb}
\usepackage{booktabs}
\usepackage{longtable}
\usepackage{array}
\usepackage{graphicx}
\usepackage{xcolor}
\usepackage{listings}
\usepackage{microtype}
\usepackage{cite}
\usepackage[hidelinks]{hyperref}
\usepackage{cleveref}

\usepackage{calc}
\usepackage{etoolbox}

\usepackage{color}
\usepackage{fancyvrb}

\DefineVerbatimEnvironment{Highlighting}{Verbatim}{commandchars=\\\{\}}

\providecommand{\ket}[1]{\left|#1\right\rangle}

\lstdefinelanguage{none}{identifierstyle=}

\title{Aicir: A Full-Stack Quantum Circuit Simulator with Ascend NPU Support}

\author{%
  Xian Lu, Xinying Li, Fei Wang, Shuai Hou,\\
  Chengkang Pan, Xin Yi, Yongmei Li \\[1ex]
  China Mobile Research Institute \\[0.5ex]
  Beijing Key Laboratory of Quantum and AI Integration Technology \\[0.5ex]
  \texttt{\{luxian,\,lixinyingyjy,\,wangfei,\,houshuai\}@chinamobile.com} \\[0.5ex]
  \texttt{\{panchengkang,\,yixin,\,liyongmei\}@chinamobile.com}
}
\date{\today}

\begin{document}
\maketitle

\begin{abstract}
Quantum computing is a promising way to study problems that are difficult for
classical methods, but current quantum hardware still faces limits in scale,
noise, and fidelity. Running quantum algorithms on physical machines can also
be costly. Quantum circuit simulators therefore remain important because they
let researchers design and test algorithms on classical computers before using
quantum hardware. Most high-performance simulators provide GPU backends, while
few offer native support for NPUs. This gap limits the computing platforms
available for quantum-algorithm research. We developed Aicir to provide a
full-stack quantum circuit simulator with a native Huawei Ascend NPU backend.
Aicir connects circuit construction, several state representations,
measurement, differentiation, variational algorithms, quantum machine
learning, and quantum architecture search through one programming model. It
also supports noise simulation, tensor-network and matrix-product-state
engines, and distributed state simulation. On the NPU, paired real tensors,
fixed-rank gate views, and hardware-specific formulas keep the tested
simulation paths on the device. The same representation lets Aicir partition a
state across $2^{p}$ NPUs while retaining reverse-mode differentiation. We
validated native execution with CPU fallback disabled and checked distributed
communication and gradients on 2, 4, and 8 NPUs. For the tested fused layered
circuits, Aicir's CPU runtime is within $0.97$--$1.28\times$ that of Qiskit Aer
and $0.76$--$1.10\times$ that of Cirq. These results place its CPU execution in
the same range as established simulators for this workload, while the NPU tests
establish correct native execution rather than CPU-to-NPU speedup.
\end{abstract}

\tableofcontents
\newpage

\hypertarget{introduction}{%
\section{Introduction}\label{introduction}}

Quantum computing offers new ways to solve problems that are difficult for
classical methods, and it has become an important platform for studying quantum
algorithms and many-body systems~\cite{preskill2018nisq}. However, present
quantum processors remain limited by noise, scale, and fidelity. Access to
physical machines is also constrained, and repeated experiments can be costly.
Researchers therefore still need classical simulators to design circuits,
check algorithms, study noise-free behaviour, and reproduce results before
using quantum hardware.

High-performance simulation increasingly relies on accelerators because an
$n$-qubit state vector contains $2^{n}$ amplitudes. Large states may also need
more than one device. Qiskit Aer, Qulacs, and QuEST provide CUDA paths
~\cite{qiskitaer,qulacs,quest}, while NVIDIA supplies state-vector and
tensor-network libraries through cuQuantum~\cite{cuquantum}. GPU acceleration
is therefore well represented in the simulator ecosystem. By contrast, we are
not aware of a full-stack quantum circuit simulator with a native and validated
NPU backend. We built Aicir to fill this gap and to give researchers another
platform on which to simulate circuits and study quantum computing.

In this work, NPU refers specifically to the Huawei Ascend 910B NPU. We built
Aicir with a native Ascend NPU backend. Aicir keeps large state
operations on the NPU and avoids extra memory passes. It also preserves
gradient paths and supports several devices. These tasks need NPU-specific code because
many standard simulation formulas use complex arithmetic and rank-$n$ tensor
views. We instead use paired real and imaginary tensors and fixed-rank gate
views. The backend intercepts operations as well, since checking the returned
tensor cannot reveal a silent CPU fallback. It executes the simulation
directly.

To the best of our knowledge, Aicir is the first full-stack quantum circuit
simulator with a native Ascend NPU
backend~\cite{tensorcircuit,mindquantum_simulator_docs,mindquantum_qaia_npu_docs,realified_tn_npu}.
We use \emph{full-stack} to mean one framework spanning circuit construction,
multiple simulation engines, measurement, differentiation, algorithmic
modules, and distributed execution. A \emph{native NPU backend} means that
simulation kernels are dispatched through an explicit NPU implementation and
validated with CPU fallback disabled. We do not include quantum-inspired
optimizers, standalone contraction methods, or execution inherited from a
generic tensor runtime without a validated simulator backend in this scope.

\hypertarget{contributions}{%
\subsection{Contributions}\label{contributions}}

\begin{enumerate}
\def\labelenumi{\arabic{enumi}.}
\item
  We built a full-stack quantum circuit simulator with a native Ascend NPU
  backend. Nineteen methods connect the numerical layer to the simulation
  engines and upper-level algorithms. Users can move from circuit construction
  to an algorithm experiment without implementing NPU kernels.
\item
  We use paired real tensors and fixed-rank gate views on the NPU. The backend
  also uses NPU-specific formulas because the NPU lacks several complex operations.
  The public programming model stays the same, but a hot backend path may
  differ from its CPU version.
\item
  We place ten architecture-search methods behind one entry point and result
  type. One multi-card mode is numerically identical to single-card execution.
  A faster mode follows a different optimization path, and we label it as such.
\item
  Aicir can partition a state vector or density matrix across $2^{p}$ devices.
  Reverse-mode autograd remains available because paired-real leaf containers
  avoid complex gradient accumulation. We checked this path on 2, 4, and 8
  cards with a machine-checkable communication contract.
\end{enumerate}

\hypertarget{scope-and-paper-organization}{%
\subsection{Scope and paper organization}\label{scope-and-paper-organization}}

In Section~2, we compare Aicir with existing quantum simulators and software
frameworks. Section~3 then introduces the system architecture, including the
backend boundary, precision policy, and validation model. Section~4 turns to
quantum-state representations, parameterized circuit construction,
measurement, and execution semantics.

Here we present the native Ascend NPU backend in Section~5 and explain its
dtype policy, numerical representation, fixed-rank gate operations, and
fallback detection.
Quantum architecture search serves as an integrated workload in Section~6. In
Section~7, Aicir extends the same backend design to distributed and
differentiable simulation across several NPUs. Section~8 collects the
variational, quantum-learning, chemistry, noise, compilation, and supporting
modules built on these layers.

The evaluation in Section~9 covers CPU performance, native NPU execution,
distributed correctness, and evaluation coverage. We retain both favourable
and unfavourable benchmark results. Section~10 discusses the remaining
performance and functional limits before outlining the roadmap. Since Aicir
currently targets circuit-level simulation on classical processors, we also
state the boundary of this scope there; its chemistry module is not an
electronic-structure solver. Finally, Section~11 concludes the work, while the
appendices provide the API reference and the information needed to reproduce
the benchmarks.

\hypertarget{related-work}{%
\section{Related work}\label{related-work}}

We compare Aicir with the systems measured in Section~9 and discuss work that
uses similar designs. The systems have different goals, so one ranking
would be misleading. For example, Qiskit Aer and Cirq are faster on most of
our measured workloads.

We use Qiskit and Qiskit Aer as reference points for circuit construction and
state-vector simulation~\cite{qiskit,qiskitaer}. Aer is our strongest measured
CPU baseline. Qiskit's
reference \texttt{Statevector} path serves a different purpose from Aer's
simulator execution path. In our four-workload table, Aicir is faster
than the reference path on GHZ, QFT, and the layered ansatz, and about 5\%
slower on the random circuit. It does not have the same advantage over Aer.

Cirq uses a different simulation method~\cite{cirq,cirq_simulator_docs}. Its
simulator enables \texttt{split\_untangled\_states} by default. It keeps
independent subsystems separate and joins them only after an entangling gate.
That factorisation may explain part of Cirq's lead in Section~9.2. It also led us to
the factored representation discussed in Section~4.1, since a product state is an MPS
with bond dimension one.

Qulacs and QuEST are compiled state-vector simulators, and QuEST also supports
distributed execution~\cite{qulacs,quest}. We did not install Qulacs in the
benchmark environment. Our local adapter also supports only
\texttt{complex128}. For this reason, we list Qulacs capabilities but report no
timing. This says nothing about its performance on another machine.

Intel-QS and NWQ-Sim support distributed state-vector simulation in HPC
settings~\cite{intelqs,nwqsim}. Their sharding work gives context for
Section~7. Our claim is narrower. Aicir keeps reverse-mode differentiation
through its partition, and we test that path separately. We make no claim
about differentiation in Intel-QS or NWQ-Sim.

The \texttt{aicir.qml.qfun} interface follows PennyLane's decorator and device
model~\cite{pennylane}. TensorCircuit connects tensor-network contraction to
JAX, TensorFlow, and PyTorch. It is the closest system here in backend choice
and differentiation. Its contract assumes established machine-learning
runtimes. We wrote our contract for a runtime that lacks some numerical
kernels~\cite{tensorcircuit}.

Yao.jl uses a typed block IR. Its reversible AD engine differentiates through
quantum evolution with less memory than naive taping~\cite{yaojl}. It addresses
a problem related to our paired-real containers, but uses a different method.

MindQuantum is the closest comparison within the broader MindSpore
ecosystem. Its \texttt{Simulator} API documents the backends
\texttt{mqvector}, \texttt{mqmatrix}, \texttt{mqmps}, \texttt{stabilizer},
\texttt{mqvector\_gpu}, and \texttt{mqvector\_cq}; no Ascend backend is
documented there. A separate NPU tutorial covers quantum-inspired algorithms
(QAIA), not the general circuit simulator. We restrict our comparison to
documented simulator support. For this reason, we do not list MindQuantum as
having a dedicated Ascend simulator backend. Aicir uses its general backend
interface for NPU execution, and we report its costs and limits directly
~\cite{mindquantum,mindquantum_simulator_docs,mindquantum_qaia_npu_docs}.

Recent work rewrites complex tensor-network contractions as real operations on
Ascend 910~\cite{realified_tn_npu}. It reports an NPU contraction method and
hardware results. It does not describe a full-stack simulator with circuit
semantics, measurement, several engines, and algorithm interfaces. We treat it
as a related numerical method for that reason.

quimb and cotengra handle tensor-network contraction and path planning. Aicir uses
\texttt{cotengra} only to choose the contraction order and slices. It executes
the contraction with its backend because an external executor would leave the
Aicir NPU and autograd path. cuQuantum supplies NVIDIA GPU kernels for
state-vector and tensor-network simulation. Aicir instead implements a native
Ascend NPU path~\cite{quimb,cotengra,cuquantum}.

We did not find a general-purpose simulator with an integrated architecture
search module in this comparison set or the cited QAS work. Those studies use
weight-shared supernets, reinforcement learning, or differentiable search for
specific tasks~\cite{du2022qas,dqas,quantumdarts}. Aicir puts ten methods behind
one interface and result type. Section~6 tests this interface. We do not count
the shared interface itself as a speed result.

Other work evaluates quantum frameworks as software. It measures circuit
construction, compilation, maintenance, and full-stack behaviour. We follow
one practical lesson from this work: we time construction and execution
separately. If we combine them, we may measure compiler cost instead of
simulation (Section~9.1)~\cite{qsebench}. We also report losses and missing
measurements.

\hypertarget{system-architecture-and-design-principles}{%
\section{System architecture and design principles}\label{system-architecture-and-design-principles}}

We designed Aicir around three goals. The programming model should remain
stable when execution moves between processors. Each backend should still be
free to use algorithms suited to its hardware. Numerical choices and unsupported
features should also be explicit, since a silent conversion or fallback can
change either the result or the device that performs the work. These goals lead
to a layered system with a narrow backend boundary.

\hypertarget{layered-system-architecture}{%
\subsection{Layered system architecture}\label{layered-system-architecture}}

Aicir has four main layers. At the top, algorithm modules such as VQE, QAOA,
and quantum architecture search consume circuits and execution primitives. The
programming layer defines typed instructions, parameters, measurement, and
classical control. Below it, the state-vector, density-matrix, tensor-network,
and MPS engines implement simulation semantics. The numerical backend forms
the bottom layer and dispatches the required work to NumPy, PyTorch, or Ascend
NPU. Distributed execution extends the simulation layer by partitioning a
state, but it does not change circuit semantics.

The layers communicate through small interfaces rather than through a shared
tensor implementation. A circuit is first reduced to typed operations. A
simulation engine then decides how those operations act on its state
representation, while the backend performs the numerical work. Algorithm
modules call \texttt{Sampler} and \texttt{Estimator} primitives, so they do not
depend directly on a particular engine or device. Appendix A lists the public
interfaces, and Section~4 describes the programming and simulation layers in
detail.

\hypertarget{backend-boundary-and-hardware-specialization}{%
\subsection{Backend boundary and hardware specialization}\label{backend-boundary-and-hardware-specialization}}

We connect \texttt{State}, \texttt{Circuit}, \texttt{Measure}, and the
algorithm modules to the numerical layer through one abstract backend. Its 19
methods cover allocation, linear algebra, state evolution, measurement,
expectation values, and conversion at the framework boundary. Adding a backend
requires an implementation of this contract, while the upper layers remain
unchanged. The same circuit can therefore run on CPU, GPU, or Ascend NPU
without changing its gate or measurement semantics.

The common interface does not require identical low-level algorithms. Aicir
keeps hardware-specific operations below the backend boundary because the
efficient formula depends on the target processor. The NumPy implementation
uses strided updates and optional in-place paths. On the NPU, Aicir uses
paired real tensors and fixed-rank gate views instead of relying on conventional
high-rank complex operations. These implementations are compared against common
semantic tests. Section~5 explains the NPU representation and execution paths,
while Section~9 gives the validation protocol.

\hypertarget{numerical-and-structural-contracts}{%
\subsection{Numerical and structural contracts}\label{numerical-and-structural-contracts}}

Each backend owns its complex dtype because one precision does not fit every
device. Aicir builds gate matrices at the widest required precision and narrows
them only at the backend boundary. This prevents an early \texttt{complex64}
construction from introducing irreversible error into a \texttt{complex128}
CPU run. It also keeps Torch readouts compatible with float32 neural-network
layers. The defaults and user-visible rules are given in Section~5.2. Precision
is part of the scientific contract because diagnostics for barren
plateaus~\cite{mccleanbarren,holmesbarren} measure gradients close to the
numerical floor.

We keep three structural contracts across the upper layers. First, circuits,
states, measurement, variational algorithms, and architecture search reach
numerical work through the backend interface rather than naming an array
framework. Second, gate metadata is the single source for matrix construction,
target axes, decomposition, export, drawing, and differentiation. Runtime
components read this record instead of maintaining separate gate lists. Third,
unsupported combinations raise at their boundary. Unbound parameters,
incompatible measurement schedules, control flow outside trajectory execution,
and invalid distributed configurations are therefore errors rather than silent
approximations. For example, the distributed path checks its measurement and
sampling constraints before allocating a state. These contracts expose backend
differences without changing the programming model. Aicir requires only NumPy
for its core, while Torch, SciPy, Matplotlib, and external contraction planners
remain optional.

\hypertarget{programming-model-and-simulation-semantics}{%
\section{Programming model and simulation semantics}\label{programming-model-and-simulation-semantics}}

Aicir uses one typed instruction representation for circuit construction,
simulation, measurement, and classical control. The same representation carries
gate metadata and symbolic parameters to the simulation engines and algorithm
modules in Sections~5--8. This section describes the resulting user model and
the semantic rules shared by all backends.

\hypertarget{quantum-state-representations}{%
\subsection{Quantum state representations}\label{quantum-state-representations}}

Aicir uses several quantum-state representations because no single form is
efficient for every circuit. Dense states provide direct access to all
amplitudes and support the broadest execution semantics. Tensor networks and
matrix product states exploit circuit or entanglement structure, while factored
and batched states address separability and repeated workloads. These
representations share the circuit semantics from this
section, but they make different time, memory, and approximation trade-offs.

\begin{longtable}[]{@{}
  >{\raggedright\arraybackslash}p{(\columnwidth - 6\tabcolsep) * \real{0.19}}
  >{\raggedright\arraybackslash}p{(\columnwidth - 6\tabcolsep) * \real{0.28}}
  >{\raggedright\arraybackslash}p{(\columnwidth - 6\tabcolsep) * \real{0.27}}
  >{\raggedright\arraybackslash}p{(\columnwidth - 6\tabcolsep) * \real{0.26}}@{}}
\toprule
Representation & Storage & Main use & Accuracy or scope \\
\midrule
\endhead
State vector & $2^{n}$ complex amplitudes & General pure-state simulation & Exact \\
Density matrix & $2^{n}\times2^{n}$ complex entries & Mixed states and noise & Exact \\
Tensor network & Gate and boundary tensors & Contraction, selected amplitudes, expectations & Exact contraction \\
MPS & Chain of rank-3 site tensors & Low-entanglement circuits & Exact without truncation; otherwise approximate \\
Factored state & Independent dense subsystem factors & Separable or locally entangled circuits & Exact; factors merge under entangling gates \\
Batched state vector & Batched real and imaginary amplitude arrays & QML and repeated parameterized circuits & Exact for its supported gate set \\
\bottomrule
\end{longtable}

The \texttt{State} class provides the default dense representation. A pure
state is stored as a backend-native column vector with shape $(2^{n},1)$, and a
density matrix has shape $(2^{n},2^{n})$. The same class distinguishes the two
forms and applies $U\lvert\psi\rangle$ or $U\rho U^{\dagger}$ as appropriate.
It also supplies probabilities, sampling, expectation values, partial traces,
purity, and conversion from a pure state to a density matrix. State-vector
storage grows as $O(2^{n})$, whereas density-matrix storage grows as
$O(4^{n})$; Aicir therefore uses the latter when mixed-state or noise semantics
require it rather than as the default pure-state representation.

The tensor-network engine represents circuit inputs, gates, and open outputs as
a network of low-rank tensors~\cite{quimb,cotengra}. Contracting every open
output produces the exact state vector, but Aicir can instead contract a single
amplitude, a selected set of amplitudes, or an expectation value without first
materialising the full state. Contraction order determines the intermediate
memory cost. Aicir may use \texttt{cotengra} for path search and slicing,
\texttt{opt\_einsum} for planning, or its own greedy planner; the selected
backend still executes each contraction, which preserves Ascend execution and
the autograd graph.

For circuits with limited entanglement, \texttt{MPSState} stores one rank-3
tensor per qubit and moves the orthogonality centre as gates are applied
~\cite{vidal2003mps}. One-qubit gates update one site. A two-qubit gate joins
neighbouring tensors, applies the operation, and separates them again with an
SVD; non-neighbouring qubits are brought together through swaps. With a
sufficient bond dimension and no discarded singular values, this procedure is
exact. Setting \texttt{max\_bond\_dim} or \texttt{cutoff} bounds memory by
discarding small singular components and therefore gives an approximate state.

\texttt{FactoredState} takes a different structural approach. It begins with
independent one-qubit dense factors and keeps them separate until a gate acts on
qubits from different factors. Aicir then joins only the affected factors and
continues exact evolution within the merged subsystem. This representation
introduces no truncation, but its advantage disappears as global entanglement
merges the factors into one dense state. It is therefore useful for separable
and locally entangled circuits rather than a general replacement for dense
simulation.

One specialised dense form serves repeated workloads.
\texttt{BatchSV} stores a batch of $B$ state vectors as separate real and
imaginary arrays of shape $(B,2^{n})$. It evolves all samples together and
supports the parameterized gate set used by Aicir's batched quantum-learning
path. Section~7 separately describes the distributed representation because its
state layout is tied to rank ownership and collective communication.

\hypertarget{circuit-representation-and-gate-metadata}{%
\subsection{Circuits, gates, and Hamiltonians}\label{circuit-representation-and-gate-metadata}}

The public construction model centres on the \texttt{Circuit},
\texttt{Operation}, \texttt{GateSpec}, and \texttt{Hamiltonian} classes. A
\texttt{Circuit} stores an ordered sequence of typed \texttt{Operation},
\texttt{Measurement}, and \texttt{ControlFlow} objects, together with the
number of qubits and an optional backend. A circuit can be created from a gate
sequence, extended one operation at a time, or composed with another circuit of
the same width. Gate factories such as \texttt{rx()} and \texttt{cx()} create
\texttt{Operation} instances, while \texttt{GateSpec} describes a reusable gate
definition. The object inspected by the runtime is therefore also the object
created by the public API. Dictionaries remain available for JSON, OpenQASM,
and older integrations, but they are a serialisation boundary rather than the
internal model. Appendix A.2 gives a minimal construction and serialisation
example.

Aicir provides the following built-in gate families:

\begin{longtable}[]{@{}
  >{\raggedright\arraybackslash}p{(\columnwidth - 2\tabcolsep) * \real{0.30}}
  >{\raggedright\arraybackslash}p{(\columnwidth - 2\tabcolsep) * \real{0.70}}@{}}
\toprule
Family & Gates \\
\midrule
\endhead
Fixed single-qubit & $I$, $X$, $Y$, $Z$, $H$, $S$, and $T$ \\
Parameterized single-qubit & $R_x$, $R_y$, $R_z$, $U_2$, and $U_3$ \\
Controlled & CX/CNOT, CY, CZ, CRX, CRY, CRZ, and Toffoli/CCNOT \\
Two- and multi-qubit & SWAP, RXX, RZZ, single excitation/Givens, and double excitation \\
User-defined & Arbitrary local unitary matrices and registered \texttt{GateSpec} definitions \\
\bottomrule
\end{longtable}

Controlled operations use one representation for their targets, controls, and
optional control states. A control may therefore trigger on
$\lvert 0\rangle$, and the same form supports several controls or targets
without introducing a new gate name. For a gate outside the built-in set, a
user may insert its local matrix through the \texttt{unitary} operation. A
reusable custom gate can instead be registered with a \texttt{GateSpec}, which
may define its name, aliases, arity, matrix constructor, decomposition, export
data, generator, and parameter-shift rule. Matrix construction, validation,
transpilation, drawing, export, and differentiation then read the same record.
For example, excitation gates select a four-term shift rule because their
generators have the \texttt{\{-1,\ 0,\ 1\}} spectrum
~\cite{schuld2019gradients}. Appendix A.3 shows the metadata interface.

Parameterized gates turn a fixed circuit layout into a family of unitaries
$U(\boldsymbol{\theta})$. This is the basic object used by variational quantum
circuits and quantum-learning models: an optimiser or a classical network
changes $\boldsymbol{\theta}$, Aicir evolves the corresponding circuit, and a
measurement returns the value used by the loss function. The built-in
parameterized families include $R_x$, $R_y$, $R_z$, $U_2$, $U_3$, controlled
rotations, $R_{XX}$, $R_{ZZ}$, and single- and double-excitation gates. Custom
gates can join the same workflow when their \texttt{GateSpec} supplies a
parameterized matrix constructor and, when needed, differentiation metadata.
For an observable $H$, this workflow defines an objective such as

\begin{equation}
  f(\boldsymbol{\theta}) =
  \langle 0|U^{\dagger}(\boldsymbol{\theta})
  H U(\boldsymbol{\theta})|0\rangle.
\end{equation}

The circuit structure stays fixed while the parameter values change. VQE and
QAOA minimise objectives of this form. A quantum-learning model can bind one
part of $\boldsymbol{\theta}$ from an input sample and train the remaining
part as model weights.

The \texttt{Parameter} class represents a named symbolic value. A circuit may
reuse one parameter in several gates or combine many parameters in one
template. \texttt{Circuit.parameters} collects the unbound values in their
first-use order, so a caller may bind a sequence in that order or use a mapping
keyed by names or \texttt{Parameter} objects. Binding normally returns a new
circuit and leaves the template unchanged. Aicir also supports explicit
in-place binding and partial binding for workflows that fill a template in
stages. Partial binding is especially useful for the input--weight split in a
quantum-learning model. Any operation that needs a numerical matrix rejects an
unbound parameter instead of inserting a default value.

Aicir accepts numerical scalars and Torch scalar tensors as gate parameters.
Torch values remain in the computation graph, which lets VQC and QML code
differentiate an expectation value through state evolution. For supported
gates, analytic parameter-shift methods use the generator and shift rule stored
with the gate. For the usual Pauli rotations, Aicir evaluates

\begin{equation}
  \frac{\partial f}{\partial \theta}
  = \frac{1}{2}\left[
  f\!\left(\theta+\frac{\pi}{2}\right)
  -f\!\left(\theta-\frac{\pi}{2}\right)\right].
\end{equation}

Excitation gates use the four-term rule noted above because their generator
has a different spectrum. This shared metadata keeps matrix construction and
differentiation tied to the same gate definition.
Section~8.1 describes the VQE, QAOA, \texttt{QFun}, and neural-network layers
built on this parameter model. Appendix A.3 gives a complete binding example.

Hamiltonians use the same typed construction model. The \texttt{Hamiltonian}
class represents
$H=\sum_i c_iP_i$ as a list of weighted \texttt{PauliString} terms. Users may
provide full-register strings, such as \texttt{("ZI", 0.3)} and
\texttt{("XX", 0.5)}, in which case Aicir infers the register width. A local
term can instead specify its qubits explicitly; for example,
\texttt{("ZZ", [0, 3], -1.0)} places a two-qubit interaction on qubits 0 and 3
of an explicitly sized Hamiltonian. Existing \texttt{PauliString} objects are
accepted as terms as well. \texttt{Observable.hamiltonian} wraps the resulting
operator for estimators and measurement, while \texttt{Observable.pauli} and
\texttt{Observable.matrix} provide typed forms for a single Pauli observable or
an arbitrary dense operator. Aicir evaluates Pauli Hamiltonians term by term
without materialising the dense matrix, as described in Section~4.4; Section~8.2
introduces the chemistry presets built on this representation.

\hypertarget{measurement-and-classical-control}{%
\subsection{Measurement, reset, and classical control}\label{measurement-and-classical-control}}

Measurement serves two different purposes in Aicir. A user may inspect an
already constructed \texttt{State}, or run a circuit whose measurements change
the trajectory. The direct state methods provide computational-basis
probabilities, sampled counts, expectation values, and reduced states.
\texttt{State.measure(shots, bit\_order)} samples the current state in the Z
basis, while \texttt{State.expectation(operator)} evaluates
$\langle\psi|O|\psi\rangle$ or $\operatorname{Tr}(\rho O)$ without sampling.
\texttt{State.partial\_trace(keep)} returns the density matrix of the selected
subsystem. These calls do not add an instruction to a circuit.

We use \texttt{Measure.run} when measurement is part of circuit execution.
Aicir distinguishes terminal readout from an in-circuit \texttt{measure()}
instruction because they have different effects:

\begin{longtable}[]{@{}
  >{\raggedright\arraybackslash}p{(\columnwidth - 2\tabcolsep) * \real{0.30}}
  >{\raggedright\arraybackslash}p{(\columnwidth - 2\tabcolsep) * \real{0.70}}@{}}
\toprule
Measurement form & Semantics \\
\midrule
\endhead
Exact execution & \texttt{shots=None} or \texttt{shots=0} runs one trajectory without terminal sampling. In-circuit measurements still execute at their positions. \\
Terminal readout & With one or more shots, Aicir performs per-qubit Z-basis measurement after the last circuit operation. It returns one $\{+1,-1\}$ eigenvalue per requested qubit and bit-string counts. \\
Joint Pauli measurement & \texttt{measure(qubits, basis=...)} without a classical target measures the joint $X$, $Y$, or $Z$ Pauli observable, projects the state, and records one eigenvalue. \\
Classical-register measurement & \texttt{measure(..., creg=...)} or \texttt{measure(..., cbits=...)} performs per-qubit Z measurement and writes bits to a named classical register. \\
Expectation value & \texttt{observables} evaluates named operators on the state before terminal readout. The result is exact for the represented state rather than estimated from terminal counts. \\
State snapshot & \texttt{snap} records the complete state after selected top-level operation indices. \\
Reduced state & \texttt{Result.reduce(R, pos)} traces out all qubits outside $R$ from the pre- or post-readout state. \\
Reset & \texttt{reset(qubits)} maps the selected subsystem to $\lvert0\rangle$. If it is entangled with the rest of the register, the remaining state can become mixed. \\
\bottomrule
\end{longtable}

Terminal readout is controlled by \texttt{measure\_qubits}. In shot mode,
\texttt{None} disables it, an empty tuple or list reads every qubit, and an
explicit list reads that subset in the supplied order. In exact mode Aicir
ignores this parameter because no terminal sample is taken. An in-circuit
measurement is different. It always runs at its place in the operation
sequence, collapses the state, and changes all later operations. The two forms
may appear in the same execution.

An in-circuit measurement without a classical target accepts one qubit or a
list of qubits, a basis in $\{X,Y,Z\}$, and an optional \texttt{id}. Aicir
records its joint eigenvalue under the operation index, and the identifier gives
the same value a stable name. When \texttt{creg} or \texttt{cbits} is present,
the basis must be Z. The first form writes measured qubits to consecutive bits
of one \texttt{ClassicalRegister}; the second maps each qubit to an explicit
\texttt{Bit}. Both forms use a trajectory-local register, so separate shots do
not share classical state.

The parameters of \texttt{Measure.run} define the execution and the data kept
in its result:

\begin{longtable}[]{@{}
  >{\raggedright\arraybackslash}p{(\columnwidth - 4\tabcolsep) * \real{0.30}}
  >{\raggedright\arraybackslash}p{(\columnwidth - 4\tabcolsep) * \real{0.16}}
  >{\raggedright\arraybackslash}p{(\columnwidth - 4\tabcolsep) * \real{0.54}}@{}}
\toprule
Parameter & Default & Meaning \\
\midrule
\endhead
\texttt{circuit} & required & Circuit to execute; its bound backend takes precedence over the backend supplied to \texttt{Measure}. \\
\texttt{shots} & 1 & Positive integer for sampled trajectories; \texttt{None} or 0 selects exact execution. Boolean and negative values are rejected. \\
\texttt{measure\_qubits} & empty & Terminal-readout selection: all qubits, an ordered subset, or \texttt{None} for no terminal readout. \\
\texttt{snap} & \texttt{None} & One operation index or a collection of indices whose post-operation states are retained. \\
\texttt{sm} & \texttt{"avg"} & Multi-trajectory aggregation mode. Only \texttt{avg} is currently implemented; \texttt{shot} and \texttt{cond} remain to be developed. \\
\texttt{seed} & \texttt{None} & Seed for in-circuit randomness, noise trajectories, and terminal sampling. \\
\texttt{initial\_state} & \texttt{None} & Initial pure state as a \texttt{State} or an array; otherwise Aicir starts from $\lvert0\cdots0\rangle$. \\
\texttt{initial\_density\_matrix} & \texttt{None} & Initial density matrix. It is mutually exclusive with \texttt{initial\_state}. \\
\texttt{observables} & \texttt{None} & Mapping from result names to operator matrices evaluated before terminal readout. \\
\texttt{return\_state} & \texttt{True} & Retains the state before terminal readout and the final state after it. Disabling this option avoids an unnecessary density-matrix aggregation when possible. \\
\texttt{return\_probabilities} & \texttt{True} & Retains the full $2^{n}$ computational-basis probability array. Disabling it saves memory when only counts or scalar results are needed. \\
\texttt{method} & \texttt{"statevector"} & Selects \texttt{statevector}, \texttt{tensor}, \texttt{mps}, or \texttt{factored} simulation. \\
\texttt{max\_bond\_dim}, \texttt{cutoff} & \texttt{None}, $10^{-10}$ & MPS bond limit and relative singular-value truncation threshold. \\
\texttt{fuse} & 0 & Maximum qubit width of fused gate blocks. Zero disables fusion; a positive setting must be at least 2. \\
\bottomrule
\end{longtable}

\texttt{Result} keeps the different outcomes separate. \texttt{state} is the
state before terminal readout, whereas \texttt{final\_state} includes terminal
collapse. \texttt{output(-1)} and \texttt{counts(-1)} address terminal
readout. An in-circuit joint measurement is addressed by its operation index or
string identifier. \texttt{classical\_counts(reg)} instead reports the final
integer value of a classical register across trajectories. The result also
contains the pre-readout probability array, named expectation values, snapshots,
backend metadata, and the measured-qubit order. In exact mode, terminal output
and counts are unavailable because terminal sampling did not occur.

Several combinations are rejected because their semantics would otherwise be
unclear. Tensor-network, MPS, and factored execution can reuse terminal readout
and observable evaluation, but they do not execute in-circuit measurement.
MPS and factored runs also start from $\lvert0\cdots0\rangle$ and do not record
snapshots. Gate fusion is limited to the state-vector method and cannot be
combined with snapshots, noise, or in-circuit measurement. Aicir raises for
these cases instead of silently dropping the requested behaviour.

Classical conditions read the trajectory-local register through
\texttt{Condition}. Aicir represents \texttt{if\_} and \texttt{while\_} as
typed \texttt{ControlFlow} nodes and evaluates their bodies after the relevant
measurement has updated the register. A while loop requires
\texttt{max\_iterations} and raises if the condition remains true beyond that
bound. These dynamic instructions run on the state-vector trajectory path.
\texttt{Circuit.unitary()} cannot represent them because measurement, reset,
and measurement-fed control do not define one unitary matrix. Appendix A.4
gives the corresponding calls.

\hypertarget{execution-primitives-and-semantic-guarantees}{%
\subsection{Execution primitives}\label{execution-primitives-and-semantic-guarantees}}

Algorithms need results rather than a particular simulator call. We therefore
place \texttt{Sampler} and \texttt{Estimator} primitives between the algorithm
and the execution method. A sampler returns probabilities or measurement
counts, while an estimator returns an observable expectation value. Both accept
one circuit or a list of circuits. They can also bind a parameter vector when
a caller supplies a template instead of an already bound circuit. For several
circuits, Aicir either broadcasts one
observable or pairs an observable list with the circuits. This interface lets
VQE, QAOA, and QML change the execution method without changing their outer
optimisation loop.

Aicir provides the following primitives:

\begin{longtable}[]{@{}
  >{\raggedright\arraybackslash}p{(\columnwidth - 4\tabcolsep) * \real{0.28}}
  >{\raggedright\arraybackslash}p{(\columnwidth - 4\tabcolsep) * \real{0.30}}
  >{\raggedright\arraybackslash}p{(\columnwidth - 4\tabcolsep) * \real{0.42}}@{}}
\toprule
Primitive & Execution method & Returned information \\
\midrule
\endhead
\texttt{StatevectorSampler} & Exact state-vector probabilities & \texttt{SampleResult} with probabilities and no sampled counts \\
\texttt{ShotSampler} & Finite-shot terminal or in-circuit measurement & Counts, probabilities, shot count, and measured qubits \\
\texttt{NoisySampler} & Noise-model execution followed by finite-shot sampling & The same sampling result with noise metadata \\
\texttt{StatevectorEstimator} & Exact state-vector expectation & Scalar \texttt{EstimateResult}; Pauli operators use the sparse path \\
\texttt{ShotEstimator} & Grouped Pauli measurements with finite shots & Energy, variance, shot count, groups, and per-term results \\
\texttt{NoisyEstimator} & Density-matrix expectation, with optional finite shots & Scalar expectation and noise-path metadata \\
\texttt{MPSEstimator} & MPS evolution with optional bond truncation & Expectation and accumulated truncation error \\
\texttt{BackendSampler}, \texttt{BackendEstimator} & User-supplied local, remote, or hardware runner & A result converted to the same sampler or estimator type \\
\bottomrule
\end{longtable}

The sampler variants separate analytic probabilities from finite-shot sampling.
The exact sampler evolves the circuit once and returns the full probability
distribution without shot noise. The shot sampler calls the measurement layer
and returns finite-shot counts for a selected qubit set or for in-circuit
measurement instructions. A noisy run follows the same interface but attaches
a \texttt{NoiseModel} before execution. Since all three return
\texttt{SampleResult}, downstream code reads the same fields for counts,
probabilities, shots, and measured qubits.

The estimator variants make the accuracy--cost choice explicit.
\texttt{StatevectorEstimator} computes an exact expectation from the simulated
pure state. \texttt{ShotEstimator} instead rotates Pauli terms into their
measurement bases, groups qubit-wise commuting terms, and allocates a finite
shot budget across those groups. It reports the estimated value, variance, and
per-term data. For noisy circuits, \texttt{NoisyEstimator} evaluates the
density-matrix state and may add finite-shot sampling. \texttt{MPSEstimator}
builds the MPS once, reuses it for the observable, and records the truncation
error caused by \texttt{max\_bond\_dim} and \texttt{cutoff}. The state-vector,
shot-based, and noisy estimators describe different physical or numerical
assumptions. MPS adds a controlled approximation when truncation is active.

Pauli observables have a separate exact path because a dense Hamiltonian wastes
memory. For
$H=\sum_j c_jP_j$, Aicir applies each $P_j$ directly to the state. A Pauli
string acts on a basis state as a bit flip and a phase,

\begin{equation}
  P\ket{b} = i^{\,n_Y}\,(-1)^{\operatorname{popcount}(b \wedge z_{\text{mask}})}\,
  \ket{b \oplus x_{\text{mask}}},
\end{equation}

so one term costs $O(2^{n})$ and never creates a
$2^{n}\times 2^{n}$ matrix. The estimator accumulates
$c_j\langle P_j\rangle$ term by term. This keeps the state-vector memory scale
instead of adding the $O(4^{n})$ storage of a dense operator. A dense
complex128 Hamiltonian needs about 4.3 GB at $n=14$ and 68 GB at $n=16$.
Before we added the sparse path, this matrix limited the variational stack near
$n\approx13$, although state-vector evolution reached $n=20$. One $n=12$
gradient evaluation fell from 520 s to 0.57 s after the change, and the
$n=14$ case takes 1.77 s. Section~8.1 describes how the variational algorithms
use this estimator.

Estimator gradients follow the same execution boundary. Given a template and
a parameter vector, \texttt{BaseEstimator.gradient} builds the objective from
the selected estimator and asks the differentiation registry for a method.
The automatic choice considers the backend, shot count, and whether noise is
active; callers may also request a method directly. Aicir returns the gradient,
the method name, and the number of objective evaluations in
\texttt{GradientResult}. This keeps the derivative tied to the execution that
produced the energy, which matters when exact, sampled, noisy, and MPS
objectives behave differently.

Hardware-aware selection uses the same rule. Given a device
\texttt{Target}, \texttt{estimator\_for\_target} first checks whether the run
needs a noise model or finite shots, then selects a density-matrix, shot-based,
or state-vector estimator supported by that target. If the execution lives in
an external service, \texttt{BackendSampler} and \texttt{BackendEstimator}
wrap a user-supplied runner and convert its output to the standard result type.
No remote system is hidden behind this adapter; the caller still owns job
submission and result retrieval. Appendix A.6 gives a short example.

\hypertarget{ascend-npu-backend}{%
\section{Native Ascend NPU backend}\label{ascend-npu-backend}}

Many high-performance quantum simulators already exist, and several provide
mature CUDA paths~\cite{qiskitaer,qulacs,quest,cuquantum}. In principle, we
could have forked one of these projects and added an Ascend device target. We
did not take this route because their accelerated execution is often tied to
CUDA libraries, device-specific kernels, internal memory layouts, and datatype
assumptions. The Ascend NPU software stack also lacks complete support for
operations on \texttt{complex64} tensors and for some complex-valued numerical
workflows used by conventional simulators. Changing the device name would
therefore not produce a native NPU backend. State evolution, measurement,
expectation values, and gradients all need formulas that match the functions
available on the NPU. The runtime must also keep tensors on the device and
detect hidden CPU fallback. Adapting an existing simulator would require
replacing much of its execution core while retaining abstractions made for
another accelerator. We therefore built Aicir from scratch instead of
adding an NPU patch to another open-source simulator. This choice gave us
control over the backend boundary from the beginning. Aicir separates circuit
semantics from numerical execution, so an NPU operation may use a different
formula without changing the upper layers. Here we describe this boundary, the
NPU representation, and the gate paths used by the native Ascend backend.

\hypertarget{the-interface}{%
\subsection{Backend contract}\label{the-interface}}

We use the \texttt{Backend} interface as the boundary between circuit semantics
and numerical execution. A state vector always has shape $(2^{n},1)$, while a
density matrix has shape $(2^{n},2^{n})$. Operations return the native tensor
type of the selected backend, and conversion to NumPy is an explicit exit from
device execution. Each backend also declares its name and complex dtype. The
upper layers can therefore work with states and operators without guessing the
array framework, device, or precision.

The contract covers four kinds of work. \emph{State construction} creates
zero-filled tensors, identity operators, and the initial
$\lvert 0\rangle^{\otimes n}$ state in the backend's native representation.
\emph{Linear algebra} provides the matrix products, tensor products, adjoints,
traces, and element-wise values needed by simulation. \emph{Quantum
operations} apply unitary evolution and evaluate inner products, reduced
states, and observable expectations. \emph{Measurement and data conversion}
turn amplitudes into probabilities and counts, import user data, and return
results to the host when requested. These groups are small enough to implement
on a new device, but they cover the numerical work used by the simulation
engines. Appendix~A.5 gives the complete list of required methods.

Aicir implements this contract in \texttt{NumpyBackend} for CPU execution,
\texttt{GPUBackend} for autograd-capable PyTorch execution on CPU or CUDA, and
\texttt{NPUBackend} for the Ascend NPU. We include quantum-level operations such
as expectation values and partial traces instead of defining the interface as
a generic tensor algebra. This matters because the best formula depends on the
device. The NPU backend can therefore replace an unsupported complex-valued
formula with paired-real arithmetic while the circuit, measurement, and
algorithm layers continue to use the same operation.

\hypertarget{ascend-limitations-and-real-valued-decomposition}{%
\subsection{Numerical constraints and design response}\label{ascend-limitations-and-real-valued-decomposition}}

Our tests on the Ascend NPU found several numerical constraints that affect
quantum simulation. The device software does not provide complete support for
operations on complex-valued tensors. Conventional formulas for state updates,
inner products, normalisation, matrix factorisation, and reverse-mode
accumulation may therefore fail even when the same work is available for real
tensors. We keep these differences inside the backend, so the circuit and
algorithm layers do not need device-specific code.

The numerical backend owns the complex dtype because precision and kernel
support differ across processors. The upper layers read this value from the
backend instead of choosing one themselves:

\begin{longtable}[]{@{}
  >{\raggedright\arraybackslash}p{(\columnwidth - 4\tabcolsep) * \real{0.33}}
  >{\raggedright\arraybackslash}p{(\columnwidth - 4\tabcolsep) * \real{0.25}}
  >{\raggedright\arraybackslash}p{(\columnwidth - 4\tabcolsep) * \real{0.42}}@{}}
\toprule
Backend & Default dtype & Reason \\
\midrule
\endhead
\texttt{NumpyBackend} & \texttt{complex128} & Correctness-first CPU execution and comparison with double-precision simulators \\
\texttt{GPUBackend} & \texttt{complex64} & Float32-oriented PyTorch execution on CPU or CUDA \\
\texttt{NPUBackend} & \texttt{complex64} (enforced) & Native execution with the numerical kernels available on the Ascend NPU \\
\bottomrule
\end{longtable}

Users may override the process-wide default for backends that support another
dtype. \texttt{NPUBackend} instead rejects unsupported precision, since silent
narrowing would hide a change in the numerical model. In the table,
\texttt{complex64} describes the precision exposed by the NPU backend. It does
not mean that every native \texttt{complex64} operation is available. Aicir
implements the required complex arithmetic with paired \texttt{float32}
tensors when a complex kernel is missing. We also construct
gate matrices at the widest required precision and cast them only at the
backend boundary. If a rotation is first built in \texttt{complex64},
converting it to \texttt{complex128} later cannot recover the lost bits. This
rule reduced the norm deviation of a nominally double-precision path from about
$10^{-8}$ to $10^{-16}$. Appendix A.5 gives the dtype configuration calls.

We designed the NPU backend around pairs of real tensors. Aicir stores a
complex value $z=x+iy$ as $(x,y)$ and evaluates the required formulas with real
arithmetic. For example, if $\psi=a+ib$ and $\varphi=c+id$, then

\begin{equation}
  \langle\psi|\varphi\rangle
  = \sum (ac + bd) + i\sum (ad - bc).
\end{equation}

The paired-real representation covers state evolution, observable
expectations, normalisation, partial traces, and matrix factorisation. For a
complex SVD, Aicir embeds the input in an equivalent real block matrix, runs a
real-valued SVD, and reconstructs the complex factors. Rank-deficient matrices
need care because the real embedding contains duplicate complex directions.
Aicir uses a fixed-count, device-side pivoted selection and performs two
orthogonalisation passes. This keeps the basis construction on the NPU and
avoids synchronising candidate norms with the host.

The representation also helps differentiation because trainable graph leaves
remain real and complex gate values are built from real parameters.
Section~7.2 explains how Aicir carries this rule into distributed gradients.
Other execution constraints concern high-rank tensor views and hidden CPU
fallback. We handle both in Section~5.3.

\hypertarget{npu-efficient-operations-and-fallback-detection}{%
\subsection{NPU execution paths}\label{npu-efficient-operations-and-fallback-detection}}

Section~5.2 described the numerical representation. Here we explain how Aicir
uses it during circuit execution. A run first creates
$\lvert0\rangle^{\otimes n}$ from real one and zero tensors and combines the
real and imaginary parts once. Gate matrices are then moved to the selected
device and cached. Aicir keeps a matrix real when all its entries are real,
because a real gate acting on a complex state needs two real matrix
multiplications. A genuinely complex gate needs four. This choice is made
inside the backend and does not change the gate seen by the circuit.

Gate application groups amplitudes that differ only on the target qubits. For
a $k$-qubit gate $U_T$, Aicir builds an index table
$I_T\in\mathbb{N}^{2^k\times2^{n-k}}$ and evaluates

\begin{equation}
  \Psi_T=\psi[I_T], \qquad
  \Psi_T'=U_T\Psi_T.
\end{equation}

The updated columns are scattered back to their original basis positions. For
small states, the common path can instead merge consecutive non-target axes
and use a grouped view whose rank is at most $2k+1$. Once an NPU run exceeds
eight qubits, Aicir selects the flat gather--multiply--scatter path. Its working
rank no longer grows with $n$, so it avoids the high-rank reshape and transpose
used by a conventional $(2,\ldots,2)$ state view. This distinction matters on
the Ascend NPU because the latter path reaches the operator-rank limit as the
circuit grows.

The NPU gate kernel works on the real and imaginary parts separately. It
gathers both parts with the same index table, multiplies them by the local gate,
and writes two real output buffers before constructing the complex result.
Aicir wraps this sequence in a custom backward rule. The backward pass gathers
the required output gradients, computes gate and state gradients with real
matrix multiplications, and uses real \texttt{scatter\_add} to restore the
state gradient. A complex tensor is constructed only after the accumulation is
complete. This avoids complex gradient addition while preserving gradients
with respect to both the state and a parameterized gate.

Measurement starts from the same representation. For an amplitude
$\psi_i=x_i+iy_i$, Aicir computes

\begin{equation}
  p_i=x_i^2+y_i^2
\end{equation}

and normalises the probability vector on the device. Exact measurement returns
this vector. A finite-shot run draws basis indices from it and forms counts
with the backend tensor operations. The state does not need to pass through a
CPU complex-number routine. Host conversion occurs only when the result layer
requests a NumPy value.

Expectation values have two NPU paths. For a general operator, Aicir evaluates
$\langle\psi|O|\psi\rangle$ with paired-real matrix products and a custom
backward rule. The state enters that rule once, so its gradient is accumulated
with real tensors before Aicir constructs the final complex gradient. Pauli
operators use the matrix-free path from Section~4.4. A Z action negates the
$\lvert1\rangle$ half of a rank-3 strided view, while an X action exchanges its
two halves. Y combines these changes with the required phase. This method
neither constructs a dense operator nor creates a $2^n$ integer index array.

Density-matrix operations use the same ideas. Aicir evaluates
$U\rho U^\dagger$ and $\operatorname{Tr}(\rho O)$ through the paired-real
matrix and trace kernels. For a partial trace, it removes one qubit at a time.
Each step reshapes the real or imaginary matrix to
$(L,2,R,L,2,R)$ and sums the two matching diagonal blocks. The working tensor
therefore has rank 6 for any $n$, instead of the rank-$2n$ view used by a
direct implementation. The reduced matrix remains on the NPU after each step.

These paths keep the large state, operator, probability, and intermediate
tensors on the selected device. Aicir uses \texttt{to\_numpy} as the explicit
exit for returned data and diagnostics. This rule is necessary but not enough
to prove native execution, because an unsupported operation may still move its
inputs to the host and copy the result back to the NPU.

We found this case in an earlier Pauli implementation. A bitwise shift used to
compute Pauli signs produced the correct output tensor, but the runtime
executed that operation on the CPU. The device--host--device trip increased an
$n=14$ Pauli-term evaluation from 3 ms on CPU to 50 ms on the NPU. Repeated
$n=16$ runs varied from 171 ms to 679 ms. After we replaced bit arithmetic
with the strided Z and X actions above, the NPU time fell to 17.9 ms and the
per-term cost stayed nearly constant with $n$.

Checking the device of the returned tensor did not reveal this fallback,
because the runtime had already copied the result back. Our strict NPU
validation therefore disables the explicit CPU fallback and intercepts the
operations issued during execution. A run fails if it requests a known
host-only path. Section~9.3 describes the hardware validation based on this
rule.

\hypertarget{npu-integration-scope}{%
\subsection{Scope and limitations}\label{npu-integration-scope}}

Users select \texttt{NPUBackend} at the execution boundary. Circuit, gate,
measurement, sampler, estimator, and algorithm interfaces do not require an
NPU-specific form. This is a shared programming model, not an identical
implementation on every processor. Circuit construction, transpilation, and
chemistry data preparation may still run on the host, while the numerical
state operations described in Section~5.3 run through the NPU backend. Our NPU
claim therefore concerns the simulation backend rather than every instruction
executed by the Python process.

The current NPU paths cover state initialisation, local gate application,
state-vector probabilities and sampling, observable expectations,
density-matrix evolution, and partial traces. Aicir also provides the
paired-real matrix and SVD operations needed by supported tensor and MPS
workflows. Variational algorithms, quantum-learning models, and architecture
search can use these operations through the same primitives as the CPU
backend. Their numerical kernels are nevertheless device-specific: CPU paths
may use direct complex arithmetic or strided updates, while the Ascend paths
use paired-real kernels, flat indexed gate application, and fixed-rank views.
We compare these implementations against common numerical oracles. This keeps
the result semantics stable, but it does not imply that every algorithm
variant, noise model, measurement schedule, or control-flow combination has a
native NPU path. A workflow is covered only when each numerical operation it
uses belongs to the validated backend surface.

NPU differentiation has a narrower boundary. Aicir supports the custom gate
and expectation gradients described in Section~5.3, together with the
algorithm paths in Section~8.1 and the distributed gradients in Section~7.2.
We do not claim arbitrary complex tensor programs or unrestricted complex
matrix differentiation on the Ascend NPU.

Hardware evidence is also per path. A successful import, a CPU fallback run,
or an output tensor located on the NPU does not establish native execution.
For that claim, we require an Ascend NPU run with explicit CPU fallback
disabled and operation interception enabled. Section~9.3 reports the paths
checked under these conditions. Capabilities without such a record may still
share the public API, but we do not present them as validated NPU execution.

\hypertarget{representative-workload-architecture-search}{%
\section{Quantum architecture search}\label{representative-workload-architecture-search}}

Automated quantum-algorithm design is a central problem in AI for Quantum
because it uses learning and search to construct circuits that would otherwise
be designed by hand. It is also becoming an important direction in quantum
computing as circuits, hardware constraints, and optimisation choices grow
more difficult to explore together~\cite{du2022qas,dqas,quantumdarts}.
Quantum architecture search (QAS) addresses this problem by choosing a circuit
structure and its parameters for a given task. To the best of our knowledge,
existing general-purpose quantum simulators do not provide QAS as a built-in
functional module. We found none in the simulators considered in Section~2 or
in our review of the cited QAS studies, which present task-specific search
methods rather than simulator modules. This gap motivated us to design QAS as
part of Aicir rather than as an external example. A search may
train thousands of candidate circuits, and each candidate is itself a
variational problem. QAS is therefore our most compute-heavy integrated
workload. It tests the simulator and algorithm layers together while also
exercising NPU execution (Section~5) and distributed memory (Section~7).

\hypertarget{one-entry-point-ten-methods}{%
\subsection{Search methods and API}\label{one-entry-point-ten-methods}}

We designed one API for methods that otherwise take different inputs and
return different records. A QAS run specifies a method, its configuration, and
the problem to be solved. Aicir normalises a Hamiltonian, target state, or
target density matrix into a common problem object before it dispatches the
request. This keeps task preparation separate from the search rule.

Aicir provides ten methods in eight families:

\begin{longtable}[]{@{}
  >{\raggedright\arraybackslash}p{(\columnwidth - 4\tabcolsep) * \real{0.24}}
  >{\raggedright\arraybackslash}p{(\columnwidth - 4\tabcolsep) * \real{0.76}}@{}}
\toprule
Method & Family \\
\midrule
\endhead
\texttt{supernet} & Weight-shared supernet, including the base,
classification, and H2 variants~\cite{du2022qas} \\
\texttt{mogvqe} & NSGA-II multi-objective topology compression~\cite{mogvqe} \\
\texttt{pporb} & Trust-region PPO with rollback~\cite{zhu2023pporb} \\
\texttt{pprdql} & Policy-reuse deep Q-learning~\cite{pprdql} \\
\texttt{crlqas} & Curriculum RL, DDQN, Adam-SPSA~\cite{crlqas} \\
\texttt{qdrats} & Differentiable architecture search~\cite{quantumdarts} \\
\texttt{dqas} & Score-function architecture gradients~\cite{dqas} \\
\texttt{vqe\_loop} & Bootstrap, low-cost screening, and fair-label VQE closed loop \\
\bottomrule
\end{longtable}

The \texttt{vqe\_loop} method plays a broader role than a single search rule.
We designed it as a full design--evaluate--learn cycle for VQE. Aicir first
builds a bootstrap pool, applies an optional low-cost structural screen, and
assigns comparable VQE labels under one fixed optimisation protocol. Later
rounds mutate labelled circuits, use a learned predictor or a low-cost fallback
to choose new candidates, and feed the new fair labels back into the next
round. The final comparison always uses the fair VQE energy. Cheap scores help
decide where to spend that evaluation budget; they do not replace it.

The \texttt{config} interface gives these methods a common way to construct
their settings. A caller may use a method-specific factory or ask
\texttt{config.create} to select one by name. Each factory supplies that
method's defaults, accepts its own search and optimisation parameters, and
rejects unknown fields before execution. The resulting typed configuration is
passed through the same \texttt{run} interface. We keep the construction API
uniform, but we do not force unrelated methods into one oversized configuration
class because their search rules need different parameters.

Each method implements the \texttt{SearchStrategy} interface and registers
under a canonical name. The dispatcher therefore stays unchanged when we add
another strategy. Most methods use PyTorch for optimisation, although the API
does not expose this implementation choice to the caller.

All strategies return \texttt{QASResult}. Its common fields record the method,
objective value, selected circuit, parameters, optimisation history, and
metadata. The result also follows the \texttt{AlgorithmResult} protocol used by
VQE, QAOA, VQD, SSVQE, and the primitives. Method-specific records remain
available through \texttt{raw}, since a single schema cannot express every
search trajectory without losing information. Appendix A.7 shows the public
interface.

\hypertarget{search-space-and-simulator-integration}{%
\subsection{Candidate evaluation and simulator integration}\label{search-space-and-simulator-integration}}

A QAS method repeatedly constructs a candidate circuit, evaluates it, and uses
the result to choose the next candidate. Aicir represents a candidate as an
\texttt{ArchitectureSpec}, which holds a \texttt{Circuit} together with its
name and metadata. Candidates may come from the built-in architecture library
or from a user-provided circuit. Search methods can then change gate choices,
connections, layers, or continuous gate parameters according to their own
rules. This representation matters because a selected architecture is already
an Aicir circuit and can be executed or fine-tuned without conversion.

Aicir supports both task-specific objectives and a common architecture
evaluator. For the latter, the evaluator scores expressibility, trainability,
noise robustness, and hardware efficiency. It then combines the four scores
with configurable non-negative weights and ranks the candidates. The active
metric in each group is explicit, so a run does not silently mix several
definitions of the same property. The current choices include distributional
distance from Haar-like states, structural or gradient-based trainability,
ideal-to-noisy sensitivity, and native-gate or topology-mapping efficiency.
Metrics that have not yet been developed cannot be selected as if they were
available.

Four training-free assessments distinguish \texttt{vqe\_loop} from a search
that ranks candidates only after full parameter optimisation. They use the
circuit structure and require no VQE iterations. Let $q$ be the number of
qubits, $L$ the number of layers, $P$ the number of parameters, $G$ the total
gate count, $G_1$ the number of single-qubit gates, $G_2$ the number of
two-qubit gates, and $D$ a circuit-depth proxy.

\emph{Structural expressibility.} We estimate whether a candidate has enough
local rotations, parameters, and entangling structure to represent a useful
set of states. The score combines rotation richness $R$, parameter density,
entanglement coverage $E_{\mathrm{ent}}$, and a final-rotation bonus $B$:
\begin{equation}
 E_{\mathrm{expr}}=\operatorname{clip}_{[0,1]}\!\left(
 0.45R+0.35\min\!\left(1,\frac{P}{3q(L+1)}\right)
 +0.20E_{\mathrm{ent}}+B\right).
\end{equation}
This is a structural proxy rather than a sampled Haar-distance measure. It is
cheap enough to apply to every bootstrap or mutation candidate. An extremely
small value can reject a circuit before VQE, while values in the lower tail of
the current candidate pool produce a soft flag.

\emph{Trainability.} Aicir next estimates whether the structure is likely to
remain manageable during optimisation. The proxy penalises depth, a large
fraction of two-qubit gates, and too many parameterised single-qubit operations
per qubit:
\begin{equation}
 E_{\mathrm{train}}=0.4e^{-D/10}
 +0.4e^{-2G_2/G}+0.2e^{-(G_1/q)/5}.
\end{equation}
The terms are clipped to the interval $[0,1]$ after they are combined. Because
depth and useful parameter density depend on the ansatz family, the loop also
compares a candidate with circuits in the same family and depth group. A low
relative value gives a soft warning. A very low score can become a hard reject
when the whole family--depth group also lies below the trainability threshold.

\emph{Entanglement coverage.} The third assessment asks how much of the
available two-qubit pattern is used. For the native supernet with a linear set
of candidate pairs, Aicir computes
\begin{equation}
 E_{\mathrm{ent}}=\min\!\left(1,\frac{G_2}{(q-1)L}\right).
\end{equation}
It favours circuits that can connect correlations across several layers and
helps distinguish a parameter-rich circuit from one whose parameters remain
mostly local. The score may add a soft flag when it falls below the configured
floor. It is also an input to the expressibility proxy. We do not call this a
noise-robustness measurement because it does not simulate a noise channel.

\emph{Hardware efficiency.} The last assessment estimates execution cost from
the native-gate ratio $G_{\mathrm{native}}/G$, depth, and the number of
two-qubit gates per qubit:
\begin{equation}
 E_{\mathrm{hw}}=0.4\frac{G_{\mathrm{native}}}{G}
 +0.3\min\!\left(1,\frac{100}{10D}\right)
 +0.3e^{-(G_2/q)/3}.
\end{equation}
This score rewards circuits that use the expected gate set and avoid needless
depth or entangling operations. The current implementation evaluates native
gates against a fixed default set, so it is a hardware-cost proxy rather than
an Ascend NPU calibration result.

Aicir combines the four values into a zero-cost feature, using equal weights by
default. In the current implementation, $E_{\mathrm{ent}}$ occupies the third
slot of the generic four-term weighting interface. Hardware efficiency affects
this composite feature, but it does not yet have an independent rejection
threshold. The loop can also enforce direct caps on $P$ and $G_2$. It assigns
each candidate one of three screening states: pass, soft flag, or hard reject.
Only hard rejects are removed before fair labeling. The composite feature is
stored for diagnosis and later analysis; it is explicitly not the final
ranking signal. Fair VQE energy still determines architecture comparisons and
provides the labels used to train the predictor in later rounds.

Candidate evaluation uses the same simulator interface as the rest of Aicir.
Simulation-based scores pass the circuit to the selected backend, while
variational methods use the simulator and differentiation layers to update
continuous parameters. Choosing \texttt{NPUBackend} therefore moves supported
state evolution and tensor operations to the Ascend NPU path described in
Section~5. The QAS API itself does not need a device-specific branch. This
connection is important because a search can evaluate far more circuits than a
single algorithm run.

Hardware-aware evaluation is still only partly connected to the device model.
The Aicir target description supplies basis gates and coupling topology to the
transpiler and target estimator, but the default QAS search space and its basic
hardware score still rely on a fixed native-gate set. The score therefore
describes this proxy unless the topology-mapping metric receives a hardware
profile. Trainability diagnostics are also available elsewhere in Aicir, but
they are not yet all connected to the QAS evaluator. Because these diagnostics
can depend on gradients near $10^{-7}$, we follow the precision policy in
Section~5.2 and use \texttt{complex128} on CPU by default. Differentiable search
under noise remains to be developed because analytic gradients through the
noise channels are not yet available.

\hypertarget{distributed-search-execution}{%
\subsection{Distributed search}\label{distributed-search-execution}}

QAS offers a natural form of task parallelism because many candidate circuits
can be evaluated independently. This differs from Section~7, where Aicir splits
one quantum state across devices. Here, each worker receives complete
candidates. The two forms can still be combined: candidates may run on
different NPUs, while a candidate that is too wide for one device may itself
use a distributed state.

When a run uses \texttt{NPUBackend} on an Ascend NPU, an initialised
\texttt{torch.distributed}, and more than one worker, supernet training enters
the distributed path automatically. Other configurations keep the single-NPU
path, so callers use the same QAS API in both cases. All workers use the same
seed because they must begin with the same weights and candidate set. Aicir
first ranks the candidates, then trains the shared model, and finally
fine-tunes the selected records.

During ranking, worker $r$ evaluates candidates $r$, $r+W$, $r+2W$, and so on,
where $W$ is the number of workers. A collective gather merges these records
and restores their global order. Each candidate is evaluated once, and the
result follows the single-device ranking. Fine-tuning uses a similar division:
each worker refines one ranked record, after which Aicir gathers the scores and
selects the global minimum. With one worker, this reduces to fine-tuning only
the highest-ranked candidate.

Shared-model training provides two modes. In \texttt{safe} mode, Aicir divides
the selection evaluations among workers, gathers their values, and chooses the
global minimum. Only worker 0 applies the gradient update, and the new
parameters are then broadcast. This mode preserves the single-device update
path, although its speedup is bounded by the number of selection evaluations
in one step. In \texttt{aggressive} mode, each worker selects, evaluates, and
differentiates its own candidate. Aicir averages the shared parameters and
gathers the selected supernet identifiers before every worker updates the same
optimiser group. This raises throughput, but it changes the optimisation
trajectory because the workers no longer train the same candidate.

The fair-label queue uses the same task-parallel idea and assigns one VQE task
to each NPU. It does not split a state. Safe mode is useful when agreement with
the single-device run matters, whereas aggressive mode favours search
throughput. We keep both choices explicit because they answer different
experimental needs.

\hypertarget{differentiable-distributed-simulation}{%
\section{Distributed quantum simulation across NPUs}\label{differentiable-distributed-simulation}}

Exact simulation becomes a memory problem before it becomes only a compute
problem. An $n$-qubit state vector contains $2^{n}$ complex amplitudes, while a
density matrix contains $2^{2n}$ entries. Adding one qubit therefore doubles
state-vector storage and quadruples density-matrix storage. A single NPU soon
becomes too small even when its compute units could process the local tensor.
Tensor networks and MPS reduce this cost by changing the representation, but
their accuracy or efficiency depends on circuit structure and entanglement.
Aicir instead distributes an exact dense state when that representation must be
preserved~\cite{quest,intelqs,nwqsim}.

We designed this path for several Ascend NPUs connected through HCCL. With
$W=2^{p}$ devices, each NPU stores $1/W$ of one state and participates in its
evolution. This lowers the per-device memory requirement by a factor of $W$ and
lets Aicir simulate states that do not fit on one NPU. It also provides more
aggregate compute and memory bandwidth. The gain is not free, because gates
that cross shard boundaries require communication. Capacity is therefore the
first advantage; speedup appears when local computation is large enough to
amortise HCCL exchanges. Distributed differentiation is built on top of the
same state and operation model rather than defining the whole model.

\hypertarget{distributed-execution-and-communication}{%
\subsection{Distributed quantum states}\label{distributed-execution-and-communication}}

Aicir runs one process on each Ascend NPU. Every process builds the same
circuit, binds to one device, and enters collective operations in the same
order. A rank identifies a storage shard. It does not represent a logical
qubit, so increasing the number of NPUs never consumes a circuit qubit or
changes the indices used by quantum gates. Appendix A.8 gives the launch
interface.

\texttt{DistState} records the global state and owns only the local shard. For
an $n$-qubit state vector, the global shape is $(2^{n},1)$ and rank $r$ stores a
contiguous block of shape $(2^{n-p},1)$. A density matrix has global shape
$(2^{n},2^{n})$. Aicir divides its rows, so each rank stores
$(2^{n-p},2^{n})$. The container also records the state kind, bit order, rank,
world size, and local and global shapes. This metadata lets the gate planner
map a logical target to either a local amplitude axis or a rank axis.

The normal execution path never reconstructs the full state on one NPU. Such a
gather would lose the memory benefit at exactly the point where it is needed.
Aicir gathers a state only through an explicit result operation. Probabilities,
expectation values, and normalisation are instead computed from local
contributions and reduced across ranks. A returned \texttt{DistState} can also
be used as the initial state of another run, so several circuit segments can be
executed without gathering between them.

Distributed construction checks the storage contract before allocation. The
world size must be a power of two, $n\geq p$, and state and gate data must use
\texttt{complex64}. Real NPU runs require HCCL; Gloo is used only for local
contract tests. The distributed backend transports Python metadata and
\texttt{float32} tensors. It separates every complex payload into real and
imaginary parts before communication and recombines them afterward. This rule
matches the paired-real NPU representation in Section~5 and avoids unsupported
complex collectives.

\hypertarget{distributed-quantum-operations}{%
\subsection{Distributed quantum operations}\label{distributed-quantum-operations}}

The cost of a distributed gate depends on the location of its target axes.
Aicir first builds a gate plan from the logical qubits and the current shard
layout. If all target axes lie inside a local shard, each NPU applies the gate
to its own amplitudes and no state data crosses the network. The operation then
has the same mathematical form as single-device gate application, but it works
on a smaller tensor.

A gate that acts on a rank axis needs amplitudes owned by another NPU. The plan
derives one or more partner ranks from the distributed target bits. Each rank
exchanges the required shard with those partners, selects the corresponding
blocks of the gate matrix, and adds the local and remote contributions. Aicir
does not gather the full state for this operation. Real and imaginary tensors
travel separately through HCCL, and the implementation can start a peer
exchange before finishing the local matrix product. This overlap reduces idle
time when communication and computation are both substantial.

State vectors and density matrices share this planning rule but use different
kernels. A state-vector gate applies $U$ to the distributed amplitude vector.
For a density matrix, Aicir evaluates $U\rho U^{\dagger}$ with a distributed
left action followed by a shard-local right action. Supported local Kraus
channels sum the distributed contributions of their component operators.
These operations keep state evolution exact within \texttt{complex64}
precision; no bond truncation or tensor-network approximation is introduced.

Global outputs follow the same local-first design. For a Pauli operator, Pauli
Hamiltonian, or explicitly local dense observable, every NPU computes its local
contribution and HCCL combines the resulting real scalars. Z-basis
probabilities come from the local amplitudes. Terminal sampling coordinates a
shared random choice and broadcasts the selected result when a global decision
is required. This makes measurement and expectation evaluation part of the
distributed simulator instead of forcing the state back to one device.

The main advantage is that most gates in a well-mapped circuit can remain
local, while only gates touching rank axes pay for communication. The planner
therefore exposes an important performance trade-off: more NPUs provide more
memory, but they also increase the number of distributed axes. Circuit layout
and gate order determine whether the added capacity also produces useful
speedup.

\hypertarget{differentiating-through-the-shards}{%
\subsection{Distributed differentiation}\label{differentiating-through-the-shards}}

Variational algorithms need derivatives of a global objective, not only a
distributed forward state. A naive implementation would gather the state or
detach communication from the computation graph. The first choice restores the
single-NPU memory limit, and the second loses gradients. Aicir instead keeps
reverse-mode differentiation across NPU shard exchanges and reductions.

When an input or parameter requires a gradient, \texttt{DistSimulator}
selects the paired-real differentiable path. Trainable pure states, density
matrices, and Stinespring factors are stored as separate real and imaginary
leaves. Aicir rejects a complex trainable leaf rather than converting it
silently, since a later backward pass could request complex gradient
accumulation that is unavailable on the Ascend NPU.

The forward pass records local gate operations and the HCCL communication that
links their shards. During backward, custom differentiation rules propagate
the adjoint through the local matrix products, repeat the required partner
exchanges, and reduce replicated parameter gradients across all NPUs. The
collective payload remains real in both directions. Thus each rank holds only a
local state shard, but a parameter receives the contribution of the global
quantum state.

Retaining every intermediate shard can still consume substantial memory.
Aicir therefore provides gradient checkpointing with no checkpointing,
automatic placement, or a user-selected interval. Checkpointing discards some
paired-real gate states during the forward pass and recomputes them in
backward. It trades extra gate execution for a smaller differentiation memory
footprint, which becomes important when the state itself already occupies most
of each NPU.

We also provide parameter-shift gradients, parameter-shift Jacobians, and
finite-difference gradients as independent checks. Parameter shift gives an
analytic gradient for supported gates~\cite{schuld2019gradients}, while finite
differences provide a numerical reference. Neither method shares the native
backward implementation. Agreement among these paths therefore tests the
distributed derivative rather than repeating the same code.

\hypertarget{what-is-refused}{%
\subsection{Scope and validation}\label{what-is-refused}}

The distributed simulator deliberately rejects operations that would break its
storage or differentiation contract. These boundaries prevent a workflow from
silently gathering the state, using a host fallback, or returning a gradient
for a non-differentiable sampling step.

\begin{longtable}[]{@{}
  >{\raggedright\arraybackslash}p{(\columnwidth - 2\tabcolsep) * \real{0.50}}
  >{\raggedright\arraybackslash}p{(\columnwidth - 2\tabcolsep) * \real{0.50}}@{}}
\toprule
Rejected & Reason \\
\midrule
\endhead
Non-power-of-two \texttt{world\_size} & Shard indexing assumes
$2^{p}$ \\
Arbitrary 2-D or column blocking & Only row sharding is implemented \\
Whole-system custom Kraus channels & No local factorisation \\
Unstructured full-system dense observables & Would require gathering the
state \\
Mid-circuit measurement, reset, control flow & Rejected before storage
allocation \\
\texttt{shots}, counts, or collapse on the differentiable path & Sampling is
not differentiable, so the pre-sampling gradient would be misleading \\
\bottomrule
\end{longtable}

The supported surface includes state vectors, row-sharded density matrices,
deterministic local Kraus noise after gates, Pauli and Pauli-Hamiltonian
observables, explicitly local dense observables, terminal Z-basis sampling,
and continued evolution from a distributed result. Forward-only and
differentiable runs share the state layout and gate planner, but the latter
uses paired-real autograd kernels.

We validate the distributed path against single-device execution and two
independent gradient oracles. The checks cover state-vector and density-matrix
evolution, gates, observables, noise, gradients, rank-to-device binding,
collective payload dtypes, HCCL use, memory invariants, and disabled CPU
fallback. Section~9.4 reports the 2-, 4-, and 8-NPU results and their numerical
errors. These are cross-NPU tests on Ascend NPUs rather than several CPU
processes presented as NPU execution.

Aicir also uses multiple NPUs for QAS candidate evaluation, fair-label queues,
and independent VQE tasks. Those modes assign a complete state to each NPU and
increase task throughput. The mode described in this section is different: it
partitions one state across NPUs and is the only mode that lowers the local
memory required by that state. Aicir may combine both levels by assigning
different candidates to device groups and distributing a wide candidate
within each group.

\hypertarget{algorithmic-and-supporting-capabilities}{%
\section{Algorithms and application workflows}\label{algorithmic-and-supporting-capabilities}}

The preceding sections described how Aicir represents, executes, searches, and
distributes quantum circuits. Here we show how those layers form complete
algorithm and application workflows. We begin with variational algorithms,
then turn to quantum learning and quantum chemistry. Supporting tools connect
these workflows to noise models, target devices, classical optimisers, and
external circuit formats. Appendix A gives the corresponding calls and short
examples.

\hypertarget{variational-algorithms-and-quantum-learning}{%
\subsection{Variational quantum algorithms}\label{variational-algorithms-and-quantum-learning}}

A variational algorithm links a parameterised circuit to a classical update
loop. Given an ansatz $U(\boldsymbol{\theta})$, an initial state
$\lvert\psi_0\rangle$, and an observable $H$, Aicir evaluates
\begin{equation}
 E(\boldsymbol{\theta})=
 \langle\psi_0\rvert U^{\dagger}(\boldsymbol{\theta})
 H U(\boldsymbol{\theta})\lvert\psi_0\rangle .
\end{equation}
An optimiser uses this value, and a gradient when required, to choose the next
parameter vector. Aicir keeps circuit construction, expectation evaluation,
differentiation, and classical optimisation behind separate interfaces. This
separation matters because the same ansatz can be tested with another backend,
estimator, or optimiser without changing the algorithm itself.

VQE is the main implementation of this workflow~\cite{peruzzo2014vqe}. A run
accepts a Hamiltonian and either an Aicir \texttt{Circuit} or a callable ansatz.
Aicir binds the current parameters, evaluates the energy, records the history,
and returns the best circuit and parameters found during optimisation. The
ansatz may come from the hardware-efficient, trapped-ion, or UCCSD builders, or
it may be supplied by the user. Adaptive ansatz construction is also available
for comparison~\cite{grimsley2019adaptvqe}. Because an ansatz is an ordinary
Aicir circuit, its gates and Hamiltonian follow the semantics introduced in
Section~4 rather than a separate VQE representation.

Energy estimation is the main connection between the algorithm and simulator.
The default exact path uses a state-vector estimator and evaluates Pauli
Hamiltonians without constructing a dense $2^n\times2^n$ matrix. A caller may
instead supply a shot-based, noisy, MPS, or external estimator. Shot and noisy
runs then keep their sampling or density-matrix semantics instead of being
silently treated as exact state-vector calculations. Parameter-shift provides
the default VQE gradient. The generator-aware rule in Section~4.2 selects the
appropriate shift formula for supported gates, so the variational loop does
not implement a second derivative system.

QAOA uses the same circuit and measurement layers for combinatorial
optimisation~\cite{farhi2014qaoa}. Aicir alternates evolution under a cost
Hamiltonian and a mixer for $p$ layers. It accepts diagonal and supported
non-diagonal Pauli costs, with first- or second-order Trotterisation when a term
decomposition is needed. The resulting circuit can be evaluated exactly or
with finite shots. Optimisation may use finite differences or the analytic
gate-level derivative path. Final sampling returns bit strings, while the
objective history records the energy used to train the angles.

Aicir also provides two ways to approximate low-lying excited states. VQD finds
states one at a time and adds overlap penalties against the states already
found~\cite{higgott2019vqd}:
\begin{equation}
 F_k(\boldsymbol{\theta})=E_k(\boldsymbol{\theta})+
 \sum_{j<k}\beta_j
 \left|\langle\phi_j\mid\psi_k(\boldsymbol{\theta})\rangle\right|^2 .
\end{equation}
SSVQE instead applies one parameterised circuit to several orthogonal reference
states and minimises a weighted sum of their energies~\cite{nakanishi2019ssvqe}.
The two methods expose the difference between sequential deflation and joint
subspace optimisation. Their current basic solvers use dense reference-state
evolution and serve as compact algorithm implementations; we do not present
them as validated distributed-NPU paths.

Backend selection affects only the operations supported by that backend.
Choosing the NPU backend moves the covered VQE or QAOA simulation kernels
to the Ascend NPU, but it does not make every estimator, noise model, or optimiser
native to the NPU. Likewise, the distributed simulator in Section~7 is an
explicit execution path rather than an automatic property of every
variational algorithm. This boundary lets us reuse the algorithm interfaces
without overstating hardware coverage.

\hypertarget{quantum-machine-learning}{%
\subsection{Quantum machine learning}\label{quantum-machine-learning}}

A quantum-learning model uses a circuit as a trainable map from classical data
to measured quantum features. For an input $x$ and trainable parameters
$\boldsymbol{\theta}$, Aicir constructs a circuit
$U(x,\boldsymbol{\theta})$ and returns an expectation value, a probability
vector, or finite-shot samples. A classical loss can consume this output and
update $\boldsymbol{\theta}$ together with the parameters of surrounding
classical layers. We use the same \texttt{Circuit}, backend, and measurement
semantics as the simulator, so a QML model does not introduce another circuit
representation.

\texttt{QFun} is the functional boundary of this workflow. It wraps a Python
function that builds a parameterised circuit and binds it to a backend,
measurement, and differentiation rule. The function may return one or several
expectation values, selected-wire probabilities, or samples. Calling its
gradient evaluates the registered rule from Section~4.2. For a probability
vector of dimension $D$ and a parameter vector of length $P$, Aicir returns the
full $D\times P$ Jacobian and reuses each shifted circuit evaluation across all
$D$ output components. Sampling remains non-differentiable and is rejected by
the gradient interface rather than assigned a surrogate derivative.

Aicir provides two PyTorch bridges because generality and throughput require
different execution paths. \texttt{QLayer} accepts any supported
\texttt{QFun}. It joins classical inputs and trainable quantum weights, calls
the quantum function in the forward pass, and uses its parameter-shift
Jacobian in backward. Gradients therefore flow to both the layer weights and a
preceding classical network. This path is flexible, but a batch is evaluated
sample by sample and parameter shift needs repeated circuit executions.

\texttt{BatchLayer} is the high-throughput path. We require a fixed circuit
template whose data parameters come first and whose trainable gates belong to
the supported single-parameter rotation family. Aicir creates one batched
state vector for all input rows, applies every template gate once to that
batch, and returns the $Z$ expectation of each qubit. Torch autograd follows
the batched tensor operations directly, so it does not run a parameter-shift
loop. The state keeps real and imaginary parts in separate tensors. This makes
the large batch operations compatible with the Ascend NPU backend described
in Section~5. Only the small real readout tensor is moved to the device holding
the PyTorch module weights before the next classical layer runs.

The batched layer can be placed between ordinary PyTorch modules. Our reference
classifier applies angle encoding and a hardware-efficient entangler, reads one
$Z$ expectation per qubit, and sends those values to a linear classification
head. The purpose is not to prescribe one QML model. It shows that data
encoding, quantum evolution, and a classical loss can share one training graph.
The current batched path trades flexibility for this throughput: it accepts a
fixed template and a restricted parameterised-gate set, and its built-in
readout is limited to per-qubit $Z$ expectations.

Aicir also supports quantum-kernel models. Given a feature map
$\lvert\Phi(x)\rangle$, it evaluates
\begin{equation}
 K(x,z)=\left|\langle\Phi(x)\mid\Phi(z)\rangle\right|^2 .
\end{equation}
For a dataset with $N$ samples, Aicir evolves the $N$ feature states as a batch
and forms the kernel matrix from real and imaginary Gram products. Pairwise
construction would repeat state preparation for $O(N^2)$ sample pairs, whereas
the batched route performs $O(N)$ state preparations followed by matrix
multiplication. This organisation is well suited to an NPU because useful work
grows with the number of samples rather than being split into many small
single-circuit calls.

Training diagnostics complete the workflow because a model that executes
correctly may still be hard to optimise. Aicir samples the actual model
parameters and measures the variance and norm of its objective gradients. It
can repeat this test across qubit counts and fit the decay of gradient variance
to detect a possible barren plateau~\cite{mccleanbarren,holmesbarren}. The QFIM
spectrum provides a second view: near-zero eigenvalues identify locally flat or
redundant parameter directions. These are diagnostics rather than proofs of
trainability. Their conclusions depend on the circuit, objective, sampling
range, and numerical precision. Torch is an optional dependency for the layer
and batched-model components.

\hypertarget{chemistry-and-supporting-modules}{%
\subsection{Quantum chemistry workflow}\label{chemistry-and-supporting-modules}}

Quantum chemistry connects classical electronic-structure preparation to a
quantum variational calculation. The workflow starts from a molecular
specification. Aicir turns it into a qubit Hamiltonian and reference data,
builds an ansatz, and then runs VQE. The first stages determine the quantum
problem. The last stages use the circuit and simulator layers described in
Sections~4 and~5. Keeping this boundary visible is important because generating
molecular integrals is different from simulating the resulting quantum circuit.

Aicir accepts a stored molecular preset or constructs a Hamiltonian from a new
molecular specification. The presets cover H$_2$, LiH, H$_2$O, NH$_3$,
N$_2$, and BeH$_2$, together with several H$_2$ mappings and reductions. Each
preset stores weighted Pauli terms, its qubit count, basis and mapping
information, and available reference-state metadata. The small presets, up to
six qubits, have dense ground-energy checks. The 12--16-qubit presets use
structural checks because dense diagonalisation would itself require the large
matrix that the simulator is intended to avoid.

For a new molecule, the optional chemistry pipeline takes its geometry, basis,
charge, spin, active-space choice, and fermion-to-qubit mapping. PySCF computes
the electronic-structure problem through Qiskit Nature. Aicir then converts the
mapped operator to its own Pauli-term Hamiltonian while preserving the chosen
bit order. The resulting object also records the electron count,
Hartree--Fock occupation, excitation indices, and provenance. Active-space
reduction happens before the mapping, so the number of simulated qubits follows
the reduced problem rather than the full orbital basis.

The reference metadata connects chemistry to circuit construction. For a
Jordan--Wigner problem, Aicir builds UCCSD directly from the qubit count,
Hartree--Fock bit string, and single- and double-excitation list. Occupied
orbitals are prepared with $X$ gates. Each excitation receives one trainable
parameter per repetition. When its orbitals are not adjacent, Aicir uses an
fSWAP network to bring them together, applies the excitation, and reverses the
network. This preserves the fermionic sign that an ordinary SWAP sequence
would miss. The ansatz builder consumes plain data and therefore remains
separate from the optional chemistry dependencies.

The same direct bridge is not claimed for every mapping. The pipeline can
produce parity- and Bravyi--Kitaev-mapped Hamiltonians, but their stored
excitation indices are structural information. They are not presented as a
mapper-correct UCCSD circuit. A user must provide an ansatz consistent with
that mapping. This check prevents a formally valid circuit from being treated
as the intended chemical ansatz.

Once the Hamiltonian and ansatz are ready, Aicir passes them to the VQE workflow
in Section~8.1 or to the VQE-oriented architecture search in Section~6. The
simulator evaluates Pauli expectations and updates circuit parameters, while
the QAS loop may compare alternative ansatz structures under its fair-label
protocol. Selecting the NPU backend moves supported circuit evolution and
expectation kernels to the Ascend NPU. Molecular drivers, integral generation,
active-space selection, and fermion-to-qubit mapping remain host-side
preparation. Aicir is therefore a quantum-circuit and algorithm framework for
chemistry workflows, not a replacement for an electronic-structure package.

\hypertarget{supporting-workflow-components}{%
\subsection{Supporting workflow components}\label{supporting-workflow-components}}

The workflows above also need noise models, circuit compilation, and data
exchange. We keep these components separate from the algorithm code because
the same circuit may be studied under different noise assumptions, compiled
for different targets, or passed to another framework. This separation also
keeps the ideal simulation path small when none of these services is needed.

Noise modelling provides the first link between an ideal algorithm and a noisy
execution setting. Aicir represents a model as a set of rules that attach
quantum channels to selected gate types. The available channels include common
Pauli, depolarizing, relaxation, damping, erasure, readout, and correlated
two-qubit errors. A rule may act after every gate or only after named gates, and
it may exclude the qubits used by the triggering operation. During simulation,
Aicir applies each matched channel to a density matrix through its Kraus
operators. Embedded Kraus matrices are cached and reused, which matters when a
channel appears after many gates or across many shots. Ion-trap parameters can
also be loaded into the same noise workflow. These models support noisy
algorithm studies and the noise-aware candidate scores discussed in Section~6;
they do not claim to reproduce every detail of a physical device.

Circuit compilation prepares an algorithm for the constraints of a target.
The target description records its qubit count, native gates, coupling graph,
and supported execution modes. Aicir first validates and normalises the circuit.
It can then remove inverse pairs, merge rotations, commute local gates, and
decompose operations into the target gate set. Layout assigns logical qubits to
physical wires, while routing inserts SWAP gates when an interaction is not
allowed by the coupling graph. The router carries the resulting permutation
forward instead of restoring the original layout after each interaction. This
usually avoids unnecessary SWAP gates, and the final mapping remains available
with the compiled circuit. The present automatic layout and routing methods are
greedy. They are useful as a practical compilation path, but they are neither
globally optimal nor noise-aware.

Data exchange closes the workflow. Aicir can save and load its circuit model in
JSON, which preserves instructions that do not have a direct standard-QASM
form. It also imports and exports supported subsets of OpenQASM 2.0 and 3.0 and
provides conversion bridges for Qiskit, PennyLane, and WuYue. OpenQASM is useful
for common gates and simple terminal measurements, whereas JSON is the safer
choice for Aicir-specific measurement semantics and control flow. Some OpenQASM
constructs, including reset, custom gates, and general classical control, are
not yet converted. We therefore treat OpenQASM support as an interoperability
layer with explicit boundaries, rather than as a complete implementation of
either language standard.

\hypertarget{evaluation}{%
\section{Execution efficiency and hardware validation}\label{evaluation}}

Here we examine Aicir's execution efficiency and verify its hardware paths. The
CPU benchmarks compare Aicir with other simulators and show how its cost changes
with circuit size. They also tell us whether the specialised execution paths
are useful in practice. Separate runs on Ascend NPUs check the NPU backend and
the distributed implementation on real hardware. These runs provide
correctness evidence because they do not include a matched CPU-to-NPU speed
comparison or a scaling study. Each one has an archived record of its software
environment, hardware configuration, and result. Appendix~B describes these
records and the conditions required to reproduce the experiments.

\hypertarget{evaluation-methodology}{%
\subsection{Evaluation scope and protocol}\label{evaluation-methodology}}

The CPU benchmarks use matched state-vector workloads, followed by a qubit
sweep and an internal Aicir ablation. This separates cross-framework execution
time from the gains due to specialised gate handling and gate fusion. Hardware
validation has a narrower purpose. On one Ascend NPU, Aicir executes
state-vector, density-matrix, and expectation workloads with CPU fallback
disabled, and independent numerical results act as references. Tests on 2, 4,
and 8 NPUs cover device binding, HCCL communication, distributed states, and
distributed differentiation. They contain no matched CPU measurements or
scaling curves, so Section~9.4 reports correctness rather than acceleration.

The CPU comparison begins with parity checks because the frameworks use
different conventions. Qiskit uses little-endian state indexing
~\cite{qiskit_bit_ordering_docs}. Qulacs defines \texttt{RX} as
$\exp(+i\theta P/2)$~\cite{qulacs_rx_docs}, while Qiskit and Aicir use
$\exp(-i\theta P/2)$. Cirq measures the \texttt{CZPowGate} exponent in units
of $\pi$~\cite{cirq_czpowgate_docs}. A mismatch would time different
computations; a circuit is timed only after its state vector agrees. GHZ and
QFT alone cannot test endianness. From
$\lvert 0\ldots0\rangle$, both outputs are invariant under bit reversal. The
random and layered circuits reveal this error. A meta-test requires at least
one such family in the parity suite. Gate counts are aligned as well. For QFT
at $n=16$, every framework executes 136 gates after conversion, which removes
differences caused by controlled-phase representations.

Build time remains separate from execution time~\cite{qsebench}; otherwise the
measurement would mix compilation with simulation. The reported statistics are
the median and IQR after warm-up. CPU thread count is pinned, and every record
includes \texttt{OMP\_NUM\_THREADS=1} and the linked BLAS library. This matters
because garbage collection, JIT compilation, and threaded BLAS can each shift
short timings.

\hypertarget{cpu-performance-and-memory}{%
\subsection{CPU execution efficiency}\label{cpu-performance-and-memory}}

The first CPU benchmark compares four circuit families at $n=18$. Every simulator uses double
precision and one CPU thread. Table~\ref{tab:cpu-families} reports execution
time only; circuit construction and transpilation are excluded.

\begin{longtable}[]{@{}llllll@{}}
\caption{Median CPU execution time at $n=18$ in milliseconds. Lower is better.}
\label{tab:cpu-families}\\
\toprule
Circuit family & Aicir & Aer & Cirq & Qiskit & PennyLane \\
\midrule
\endhead
GHZ & 7.79 & 11.6 & \textbf{4.36} & 13.9 & 6.79 \\
QFT & 68.8 & 40.2 & \textbf{16.5} & 155 & 112 \\
Random & 80.3 & \textbf{20.8} & 31.9 & 76.4 & 55.5 \\
Layered ansatz & 148 & \textbf{45.8} & 53.5 & 157 & 93.1 \\
\bottomrule
\end{longtable}

The result depends strongly on circuit structure. Aicir is $1.49\times$ faster
than Aer on GHZ, but $3.9\times$ slower on the random circuit. Cirq has the
lowest time on GHZ and QFT, while Aer leads on the two deeper families. Against
Qiskit's reference \texttt{Statevector}, Aicir is faster on GHZ, QFT, and the
layered ansatz; the random-circuit difference is about $5\%$. A single value of
$n$ cannot show scaling, so the layered ansatz is used for a separate sweep.
The simulators also take different execution paths. Cirq enables
\texttt{split\_untangled\_states} by default and joins factors only when a gate
entangles them~\cite{cirq_simulator_docs}. Disabling this option raises its time
by $1.14\times$ on GHZ and $1.97\times$ on the layered ansatz. Aicir instead
reduces state movement with dedicated diagonal and permutation paths. It may
reuse an input buffer when no retained state refers to it. Gate fusion then
joins consecutive operations without changing their order. A unitary check
against the unfused circuit precedes every fused timing.

Table~\ref{tab:cpu-scaling-unfused} follows the unfused specialised path from
$n=16$ to $n=22$. The layered ansatz resembles the repeated parameterised
layers used by variational algorithms. All four simulators were timed in the
same run.

\begin{longtable}[]{@{}lllll@{}}
\caption{Median CPU execution time for the unfused layered ansatz in seconds.}
\label{tab:cpu-scaling-unfused}\\
\toprule
$n$ & Aicir & Cirq & Aer & PennyLane \\
\midrule
\endhead
16 & 0.0318 & 0.0152 & \textbf{0.0120} & 0.0183 \\
18 & 0.145 & 0.0601 & \textbf{0.0469} & 0.106 \\
20 & 0.732 & 0.307 & \textbf{0.235} & 4.22 \\
22 & 3.90 & 1.26 & \textbf{1.08} & 18.7 \\
\bottomrule
\end{longtable}

Without fusion, Aicir takes $1.7\times$ to $3.9\times$ as long as Cirq and
$1.7\times$ to $3.6\times$ as long as Aer. The comparison with PennyLane
changes at larger sizes. Aicir is slower at $n=16$ and $n=18$, then becomes
$5.8\times$ and $4.8\times$ faster at $n=20$ and $n=22$. This crossover is
specific to the tested circuit and backend settings. Fusion gives the largest
gain in the Aicir ablation. As shown in
Table~\ref{tab:cpu-fusion-ablation}, it reduces the operation count by about
$5.1\times$. Execution becomes $2.60\times$ to $3.35\times$ faster than the
unfused specialised path and $3.58\times$ to $5.36\times$ faster than the
generic gate path.

\begin{longtable}[]{@{}lllll@{}}
\caption{Gate-fusion ablation on the layered ansatz. The last two columns report
speedup over the unfused specialised and generic paths.}
\label{tab:cpu-fusion-ablation}\\
\toprule
$n$ & Gates & Reduction & Specialised & Generic \\
\midrule
\endhead
16 & $188 \to 36$ & $5.22\times$ & $2.60\times$ & $3.58\times$ \\
18 & $212 \to 42$ & $5.05\times$ & \textbf{$3.35\times$} & \textbf{$5.36\times$} \\
20 & $236 \to 46$ & $5.13\times$ & $2.70\times$ & $4.74\times$ \\
22 & $260 \to 51$ & $5.10\times$ & $2.65\times$ & $4.33\times$ \\
\bottomrule
\end{longtable}

The fused and unfused states have overlap $1.0000\pm 10^{-8}$ at every tested
size. Gate-count reduction does not translate directly into the same runtime
gain because a fused $k$-qubit block uses a dense $2^{k}\times 2^{k}$ GEMM.
Many of the replaced operations are cheaper diagonal or permutation gates. The
fused Aicir path is then compared with the other simulators in
Table~\ref{tab:cpu-scaling-fused}. Aer enables its own fusion at these sizes,
so both Aicir and Aer use fusion. The Aer time includes amplitude reordering for
the common bit order. The Aicir time includes construction of the full
$2^{n}$ probability array. These adapter costs remain in the measurements.

\begin{longtable}[]{@{}lllll@{}}
\caption{Median CPU execution time for the fused layered ansatz in seconds.}
\label{tab:cpu-scaling-fused}\\
\toprule
$n$ & Aicir & Cirq & Aer & PennyLane \\
\midrule
\endhead
16 & 0.0123 & 0.0161 & \textbf{0.0121} & 0.0191 \\
18 & \textbf{0.0477} & 0.0630 & 0.0491 & 0.1065 \\
20 & 0.2647 & 0.3015 & \textbf{0.2338} & 4.2650 \\
22 & 1.3963 & 1.2690 & \textbf{1.0898} & 18.9863 \\
\bottomrule
\end{longtable}

With fusion, the Aicir-to-Cirq time ratio ranges from $0.76$ to $1.10$. The
ratio to Aer ranges from $0.97$ to $1.28$. Aicir is faster than PennyLane at
every tested size, by $1.6\times$ to $16.1\times$. Differences within the
measured run-to-run drift, about $8\%$ at small $n$ and $1\%$ at large $n$,
do not establish a stable ranking. Peak allocation at $n=18$ was also recorded
with \texttt{tracemalloc}. Relative
to the $2^{n}\cdot16\,\mathrm{B}$ state size, the values are $3.00\times$ for
Aicir, $4.00\times$ for Qiskit, $5.01\times$ for Cirq, $1.00\times$ for Aer,
and $2.51\times$ for PennyLane. These numbers are not comparable across
frameworks because \texttt{tracemalloc} cannot see Aer's C++ allocation. The
Aicir value comes from the unfused path and is used only as an internal memory
reference.

\hypertarget{axis-e-ascend-npu}{%
\subsection{Ascend NPU validation}\label{axis-e-ascend-npu}}

The single-card tests check whether Aicir can execute representative workloads
on an Ascend NPU without moving unsupported work to the host. Both workloads
ran on \texttt{npu:0} with \texttt{fallback\_to\_cpu=false}. Independent CPU
results provide the numerical reference. The sparse Pauli workload passed all
eight cases. At \texttt{complex64} precision, the errors were
$1.7 \times 10^{-8}$ for the state vector,
$2.2 \times 10^{-8}$ for the Y string, and $1.7 \times 10^{-8}$ for the
density matrix. Runtime rose from 17.87 ms at $n=14$ to 32.34 ms at $n=20$,
while the amplitude count grew by $64\times$.

The factored engine passed all six cases. It ran a separable 28-qubit circuit in
21.42 ms and kept 28 width-1 factors. A dense state would need 2.15 GB, but
Aicir computed the expectation without building it. The per-operation
measurements explain why these runs are validation rather than an acceleration
result. A Pauli term takes 0.66, 0.66, 0.69, and 0.83 ms
at $n=14,16,18,20$, while a factored gate takes 0.765 ms at $n=28$. The
workloads move very different amounts of data, yet their times are close.
Kernel launch cost therefore dominates at these sizes. A matched CPU baseline
and larger workloads are still required before drawing a speedup conclusion.

\hypertarget{coverage-and-unmeasured-dimensions}{%
\subsection{Distributed NPU validation}\label{coverage-and-unmeasured-dimensions}}

The distributed tests use 2, 4, and 8 Ascend NPUs with CANN 8.0.RC3. CPU
fallback is disabled. All thirteen test groups passed. The records confirm
rank-to-device binding, HCCL use, and real communication payloads. They also
show that the collective path does not pass complex tensors directly to HCCL.
This evidence checks the distributed state and communication contract.
Distributed differentiation was compared with parameter-shift and
finite-difference references. Table~\ref{tab:distributed-gradient-errors}
reports the largest observed errors.

\begin{longtable}[]{@{}ll@{}}
\caption{Errors in distributed gradient validation. Here $W$ is the number of
Ascend NPUs.}
\label{tab:distributed-gradient-errors}\\
\toprule
Comparison & Error \\
\midrule
\endhead
Native vs parameter-shift & $0.0$ (at most $2.4 \times 10^{-7}$) \\
Native vs finite difference, $W=2$ & $2.94 \times 10^{-5}$ \\
Native vs finite difference, $W=4$ & $3.54 \times 10^{-5}$ \\
Native vs finite difference, $W=8$ & $6.46 \times 10^{-5}$ \\
\bottomrule
\end{longtable}

The native and parameter-shift gradients agree to at most
$2.4 \times 10^{-7}$~\cite{schuld2019gradients}. Finite-difference errors stay
below $6.5 \times 10^{-5}$ for every tested world size. These results cover
Aicir only. No strong- or weak-scaling experiment was run, so the data do not
support a claim about multi-NPU speedup or parallel efficiency.

\hypertarget{limitations-and-roadmap}{%
\section{Limitations and future work}\label{limitations-and-roadmap}}

Here we discuss the functions that remain outside Aicir's current scope and
the directions in which we plan to extend the simulator.
\label{functional-gaps}
\label{roadmap}

Aicir targets classical accelerators and does not connect to a QPU. Its
compiler stops at gates and device topology, so it neither generates waveforms
nor solves continuous-time Hamiltonians. The channel models in Section~8.4 are
useful for circuit-level noise studies, but their parameters are not calibrated
to a specific quantum processor. We therefore do not claim device-level noise
accuracy. Aicir also supports classical control only on the
measurement-trajectory path. The unitary, tensor-network, and factored engines
reject \texttt{if\_} and \texttt{while\_} because these operations branch on
measurement outcomes. Physical-system simulation will need continuous-time
Hamiltonian solvers, pulse evolution, calibrated noise, and open-system
dynamics. We plan to place these functions in a separate layer because they
describe physical evolution below the circuit model.

The QAS module still relies on proxy hardware scores and lacks some
trainability, device-cost, and noise-aware metrics, as discussed in
Section~6.3. Passing \texttt{Target} directly into architecture search would
let its evaluators use the intended device constraints instead of a separate
description. We will also extend the metrics so that candidate ranking can
account for trainability, noise, and execution cost more directly.

Visualisation remains a host-side function and is not included in NPU timing.
Aicir supports subsets of OpenQASM 2.0 and 3.0, but a complete round trip for
\texttt{reset}, custom gates, and general classical control is still needed.

A compiled high-performance backend, similar in role to Qiskit Aer, is one of
our next steps. The existing NumPy engine uses buffer reuse, specialised gate
paths, and fusion. A compiled implementation can go further by reducing
dispatch overhead, reusing memory more closely, and providing faster batched
and density-matrix kernels without changing the circuit semantics. We will
also develop the factored engine for circuits whose entanglement remains
limited. Matched variational benchmarks can then compare complete energy and
gradient paths across frameworks rather than isolated circuit calls.

The algorithm layer needs broader and more complete workflows. For VQC and
QML, we plan to connect model construction, training, evaluation, and model
reuse through one interface. The chemistry module will gain more ways to build
Hamiltonians and ansatz circuits, together with the post-processing needed to
interpret the result. Optimisation requires a wider set of problem encodings
and classical optimisers. QRC raises a different set of needs because temporal
data must pass through encoding, reservoir evolution, observable extraction,
and readout training. Its interface should connect these stages without making
users assemble the whole pipeline by hand. Although these modules serve
different tasks, they can share Aicir's circuit, backend, and differentiation
layers. This also keeps their execution choices consistent across CPU, NPU,
and distributed settings.

Distributed quantum algorithms require more than splitting one simulated
state across NPUs. Aicir already partitions simulation data, but it does not
yet represent the protocols of a distributed quantum algorithm. We therefore
plan to add a distributed quantum computing module with explicit inter-node
communication, distributed circuit composition, entanglement resources, and
coordination for variational optimisation and quantum learning. Its evaluation
will report communication separately from local simulation. The
\texttt{aicir.wireless} subpackage and detailed QPU control remain longer-term
extensions.

\hypertarget{conclusion}{%
\section{Conclusion}\label{conclusion}}

We built Aicir as a full-stack quantum circuit simulator for Ascend NPUs. A
common numerical interface connects circuit construction, simulation, and
algorithm modules to a native NPU backend. Since standard complex formulas do
not map directly to all available kernels, the backend represents complex
values with paired real tensors and applies gates through fixed-rank views.
The same representation supports state partitioning across $2^{p}$ NPUs while
retaining reverse-mode differentiation. Above the simulator, Aicir brings ten
QAS methods into one interface and makes architecture search a second line of
work rather than an external example.

The experiments establish native NPU execution and check gradients and
communication on 2, 4, and 8 NPUs. They do not establish CPU-to-NPU speedup or
multi-NPU scaling. We keep those conclusions separate from the CPU comparison,
where specialised gate paths and fusion reduce the cost of several workloads.
Hardware claims are backed by operation-level validation because a returned
tensor alone cannot rule out CPU fallback.

The next stage is to test workloads large enough to show where NPU execution
becomes useful, then measure multi-NPU scaling. A compiled backend,
physical-system simulation, fuller algorithm modules, and distributed quantum
computing will extend the present circuit-level system.

Aicir is open source at
\url{https://gitee.com/Quntelligence/quantum_frame}.

\section*{Acknowledgments}
This work was supported by the Beijing Key Laboratory of Quantum and AI Integration Technology.

\bibliographystyle{IEEEtran}
\bibliography{refs}

\appendix
\section{API reference}
\label{app:api}

We discuss design and behaviour in the main text. Here we list the public calls
and give short examples.

\subsection{Installation}

\begin{lstlisting}[language=bash]
pip install -e ".[all]"     # torch / viz / sci / tn / chem / dev extras
\end{lstlisting}

Aicir requires only NumPy. PyTorch, SciPy, and Matplotlib are optional.
Components that need an optional package are unavailable when it is not
installed.

\subsection{Circuit construction and serialization}

\begin{lstlisting}[language=Python]
import aicir as A
from aicir import Circuit

circuit = Circuit(A.hadamard(0), A.cnot(1, [0]), n_qubits=2)
operation = circuit.operations[0]       # typed Operation
gate_dicts = circuit.to_gate_dicts()    # interoperability representation
\end{lstlisting}

Aicir takes a target before the control qubits. A gate may also set
\texttt{control\_states}. Typed instructions are the internal form. We use
dictionaries only for serialization and interoperability.

\subsection{Gate metadata and symbolic parameters}

\begin{lstlisting}[language=Python]
import aicir as A
from aicir import Circuit, Parameter
from aicir.gates import get_gate_spec, gate_shift_rule

get_gate_spec("rx").generator
gate_shift_rule("single_excitation")

theta = Parameter("theta")
template = Circuit(A.rx(theta, 0), n_qubits=1)
bound = template.bind_parameters({theta: 0.5})
\end{lstlisting}

Binding returns a new circuit by default. Users can request partial or in-place
binding with \texttt{allow\_partial=True} or \texttt{inplace=True}.

\subsection{Measurement and simulation engines}

\begin{lstlisting}[language=Python]
from aicir import Measure, NumpyBackend

measure = Measure(backend=NumpyBackend())
exact = measure.run(circuit, shots=None)
sampled = measure.run(circuit, shots=1024, measure_qubits=[0, 1])

marked_circuit = circuit + Circuit(
    A.measure([0, 1], basis="Z", id="m0"),
    n_qubits=circuit.n_qubits,
)
marked = measure.run(marked_circuit, shots=1024,
                     measure_qubits=[0, 1])
marked.output("m0")   # in-circuit joint-Pauli outcomes
marked.counts(-1)     # terminal bit-string counts

dense = measure.run(circuit, method="statevector")
tensor = measure.run(circuit, method="tensor")
mps = measure.run(circuit, method="mps", max_bond_dim=16,
                  cutoff=1e-10)
factored = measure.run(circuit, method="factored")
\end{lstlisting}

In-circuit measurement and terminal readout may be used in the same run. With
\texttt{shots=None} or 0, Aicir executes no terminal readout even when
\texttt{measure\_qubits} is supplied.

\subsection{Backends and precision}

The \texttt{Backend} contract contains 19 required members. They are grouped
below by their role in simulation:

\begin{verbatim}
Metadata:
  name

State construction:
  zeros  eye  zeros_state

Linear algebra:
  matmul  kron  dagger  trace  real  abs_sq

Quantum operations:
  apply_unitary  inner_product  partial_trace
  expectation_sv  expectation_dm

Measurement and data conversion:
  measure_probs  sample  cast  to_numpy
\end{verbatim}

\begin{lstlisting}[language=Python]
import numpy as np
import aicir as A
from aicir import NumpyBackend, GPUBackend, NPUBackend

cpu = NumpyBackend()             # complex128 by default
gpu = GPUBackend(device="cuda") # complex64 by default
npu = NPUBackend()               # complex64 enforced

A.set_default_dtype(np.complex128)
A.reset_default_dtype()
\end{lstlisting}

The override applies to the process. The NPU backend raises on unsupported
precision, so it never narrows a dtype silently.

\subsection{Execution primitives}

\begin{lstlisting}[language=Python]
from aicir.primitives import StatevectorEstimator
from aicir.vqc import BasicVQE

estimator = StatevectorEstimator()
vqe = BasicVQE(..., energy_estimator=estimator)
\end{lstlisting}

State-vector, shot-based, noisy, and MPS variants use the same sampler and
estimator roles. A backend implementation can connect an external runner.

\subsection{Variational methods and architecture search}

\begin{lstlisting}[language=Python]
from aicir.vqc import run_vqe
from aicir.ansatze import hea
from aicir import qas

energy = run_vqe(hea(n_qubits=4, layers=2), hamiltonian)
search = qas.run("supernet", problem=hamiltonian)
search.value, search.circuit, search.parameters
search.history, search.metadata, search.raw
\end{lstlisting}

All architecture-search methods use the same entry point and result fields.
Aicir keeps method-specific output in \texttt{raw}.

\subsection{Distributed simulation}

\begin{lstlisting}[language=Python]
from aicir.distributed import DistSimulator

sim = DistSimulator.from_env()  # world_size=2^p and n_qubits >= p
result = sim.run(circuit, observables=observables, layout=layout)
local_state = result.state      # remains sharded unless explicitly gathered
\end{lstlisting}

The distributed API also exposes backend, sharded-state, and result types. A
user must request a full state or probability vector explicitly.

\section{Benchmark reproducibility}
\label{app:reproduction}

Each result in Section~9 is linked to an experiment record, which we archive
only when it contains the conditions needed to interpret the run. Before a CPU
timing is accepted, the benchmark compares the resulting state vectors and
emits no timing if they differ. These runs use double precision and one CPU
thread with \texttt{OMP\_NUM\_THREADS=1}; their records include the framework
version, linked BLAS implementation, warm-up policy, timing samples, and
verdict. The recorded framework versions are Qiskit 2.4.2, Qiskit Aer 0.17.2,
and Cirq 1.7.0. Ascend validation instead runs in strict hardware mode with
CANN 8.0.RC3, \texttt{device=npu:0}, and
\texttt{fallback\_to\_cpu=false}. A run is rejected when no NPU is available,
and its record adds the Ascend model, device binding, fallback setting, and
HCCL checks. The test suite reported 2396 passed tests and 8 skipped tests.
Together, these records trace each reported value to its configuration and are
included in the supplementary material. They remain separate from the Aicir
package, which contains the simulator and its public interfaces.

\end{document}